\documentclass[aps,pre,amsmath,amssymb,showpacs,twocolumn]{revtex4-1}
\usepackage{psfrag,tabls,bm,bbm,color,graphicx,amsopn,dsfont}

\usepackage[utf8]{inputenc}

\newcommand{\vect}[1]{\boldsymbol{#1}}

\usepackage{epstopdf}		
\begin{document}

\title{ Quantum corrections of the truncated Wigner approximation applied to an exciton transport model}
\author{Anton Ivanov}
\email[]{anton.ivanov@physik.uni-freiburg.de}
\author{Heinz-Peter Breuer}
\address{Physikalisches Institut, Universit\"at Freiburg, Herrmann-Herder-Stra{\ss}e 3, D-79104 Freiburg, Germany}

\date{\today}

\begin{abstract}
We modify the path integral representation of exciton transport in open quantum systems such that an exact description of the quantum fluctuations around the classical evolution of the system is possible. As a consequence, the time evolution of the system observables is obtained by calculating the average of a stochastic difference equation which is weighted with a product of pseudo-probability density functions. From the exact equation of motion one can clearly identify the terms that are also present if we apply the truncated Wigner approximation. This description of the problem is used as a basis for the derivation of a new approximation, whose validity goes beyond the truncated Wigner approximation. To demonstrate this we apply the formalism to a donor-acceptor transport model.
\end{abstract}

\pacs{05.60.Gg,03.65.Yz,02.70.Ss}
\keywords{exciton transport, truncated Wigner approximation}

\maketitle


%

\section{Introduction}

	The exciton transport in large networks that are coupled to some environment is still a topic that attracts attention because of the experimental results from the last decade. Ultrafast non-linear spectroscopy used to probe the energy transfer in FMO complexes has revealed strong vibrational coherences of multiple pigments, that persist at much larger times scales than initially expected \cite{Brixner_2005,Engel_2007,Fujihashi_2015_n}. This has opened a discussion on the question whether or not these coherences have a significant impact on the efficiency of transport processes.
	
	To study theoretically possible quantum effects that are responsible for efficient transport properties of a given network, one has to pay special attention to the way how the effect of the environment on the system is modeled. The chosen description has to be simple enough to lead to an efficient and fast scheme for the simulation of different realizations of the network and it has to be able to take into account the most important effects of the environment on the system. For example, one can construct a time local equation for the system density matrix where the effects of the environment, like dephasing, are described by Lindblad operators. To take into account the non-Markovian effects arising from the interaction between the system and the environment we have to model the latter as an infinitely large set of harmonic oscillators, that are linearly coupled to the system. These new degrees of freedom can be integrated out exactly from the problem, which makes the effective action of the system non-quadratic in the system fields and also time non-local. This description of the problem requires the use of advanced numerical methods like the Hierarchical equations of motion (HEOM) technique \cite{Tanimura_1989_n, Tanimura_1990_n, Ishizaki_2005_n, Tanimura_2006_n, Zhu_2013_n, Moix_2013_n}, the quasi-adiabatic propagator path integral method \cite{Makri_1995_n, Makri_1995_nn, Thorwart_2000_n}, the multilayer multiconfiguration time-dependent Hartree approach \cite{Wang_2006_n}, the density matrix renormalization group method \cite{Chin_2010_n} and other methods \cite{muhlbacher_2012,Segal_2010,Weiss_2008}, which in most cases are numerically expensive.
	
	We are interested in approximate methods, where one can find a good compromise between the accuracy of the results and the applied numerical effort. One of the methods, whose complexity scales linearly with the size of the network and thus can be preferred for the case of working with large networks, is the truncated Wigner approximation (TWA). In this approximation one neglects some of the non-quadratic terms in the effective action. This approximation is often applied to describe the dynamics of ultracold atomic gases  \cite{Polkovnikov_2003_b,Isella_2005,Isella_2006,Scott_2006,Martin_2010,Norrie_2006,Polkovnikov_2004,Barnett_2011} and to explain some experimental results \cite{Gross_2011}. Its validity for closed systems at short time scales was studied in \cite{Polkovnikov_2003}. Its ability to reproduce the non-Gaussian statistics of the single mode anharmonic oscillator was shown in \cite{Corney_2015}, and its applicability to the calculation of multitime correlation functions was studied in \cite{Berg_2009}. Relying on these results the application of the approach to open quantum systems was also made in \cite{Ivanov_2013}, assuming that the effect of the environment, expressed mainly in the form of additional noise, will wash out all quantum scattering processes which occur at long times.
	
	In the following, a different two-step approximation that extends beyond the TWA, will be derived for the special case of working with a Frenkel exciton Hamiltonian. In the first step, similarly to the TWA, some of the variables in the corresponding path integral representation of the problem will be analytically integrated out, the integration of the remaining set of variables in the path integral will be equivalent to solving an equation of motion of these variables. We will do the analytic integration without neglecting any terms from the action, but the price that we have to pay is the introduction of a set of pseudo-probability density functions in the path integral representation of the problem. From the obtained equations of motion we can clearly identify the contributions, that were also present in the TWA as well as the new terms. Since this representation of the problem is numerically expensive, in the second step we adopt a numerically applicable approximation that extends beyond the TWA for the case of preparing the system initially in the single exciton manifold. We will refer to it as the corrected truncated Wigner approximation (CTWA). The ability of both approximations to reproduce the exact system dynamics will be studied and we will see that both approximations reproduce exactly the short time behavior of the system. In the limit of weak couplings the CTWA can also extend this range to longer time scales. 
	
	The paper is structured as follows. In Sec. \ref{subsec:Model_system} we introduce the system of interest and its path integral representation, in Sec. \ref{subsec:Wigner_formalism} the Wigner formalism is briefly described, in Sec. \ref{subsec:Mapping_proc} a mapping procedure is presented, which makes the action local in time at the cost of introducing integrals over new variables. This step is necessary if we want to integrate out exactly the fields of the path integral representation of the system. In Sec. \ref{subsec:TWA_new} the TWA is briefly explained and the corresponding stochastic equation for the semi-classical description of our problem is derived. It will be used later for comparison with the equation obtained from CTWA. An exact stochastic equation, which takes all terms into account, that have been neglected in the TWA is derived in Sec. \ref{sec:Stochastic_equation_with_all_quantum_corrections} and a numerically applicable approximation of this equation is derived in Sec. \ref{sec:TWAQC}. The accuracy of the new approximation is tested in Sec. \ref{sec:Examples} for a donor-acceptor transport model. The paper concludes with a summary in Sec. \ref{sec:Conclusions}.

\section{Theory}
\label{sec:Theory}
\subsection{ Model system}
\label{subsec:Model_system}
We consider a system (S) composed of $\mathcal{N}$ sites, each of them being linearly coupled to a bath (B) of harmonic oscillators. The corresponding Hamiltonian is given by:
\begingroup
\allowdisplaybreaks
\begin{align}
\allowdisplaybreaks
\hat{H} 					& = \hat{H}^{}_{S} + \hat{H}^{}_{B,SB}, \label{eq:Def_H_tot} \\
\hat{H}^{}_{S} 				& = {   \sum\limits^{\mathcal{N}}_{n,n'=1} } h^{}_{nn'} \hat{a}^{\dagger}_{n} \hat{a}^{}_{n'} , \label{eq:Def_HS}\\
\hat{H}^{}_{B,SB}			& = {  \sum\limits^{\mathcal{N}}_{n=1} \sum\limits^{ }_{k} }
				\big[
				\omega^{}_{nk} \hat{b}^{\dagger}_{nk} \hat{b}^{ }_{nk} + 
				\lambda^{}_{nk} \big( \hat{b}^{\dagger}_{nk} + \hat{b}^{}_{nk} \big)\hat{a}^{\dagger}_{n} \hat{a}^{}_{n}
				\big],\label{eq:Def_H_BSB}
\end{align}
\endgroup
where $\hat{a}^{\dagger}_{n},\hat{b}^{\dagger}_{nk}$ create an excitation at the $n$-th site of energy $h^{}_{nn}$ or in the $k$-th bath mode of energy $\omega^{}_{nk}$ that is coupled to the $n$-th site. The initial state of the total system is of the form:
\begin{align}
\hat{\rho}^{}_{tot} & = \hat{\rho}^{}_S \otimes \hat{\rho}^{eq}_B, \\
\hat{\rho}^{eq}_{B}	& = \prod\limits^{ \mathcal{N} }_{n=1} \prod\limits^{}_{k} \exp [-\beta \omega_{nk} \hat{b}^{\dagger}_{nk} \hat{b}^{}_{nk} ]/N,
\end{align}
where $\hat{\rho}^{}_S$ and $\hat{\rho}^{eq}_B$ are the density matrix of the system and the equilibrium density matrix of the bath, $\beta = 1/(k_B T)$, $T$ is the temperature and $N$ is a normalization constant. We assume that the initial state of the system $\hat{\rho}_S$ lies in the single exciton subspace. Since the Hamiltonian \eqref{eq:Def_H_tot} conserves the total number of excitations it reduces to the Frenkel exciton Hamiltonian \cite{Fleming_2002}. Since we work in the single exciton subspace we have the freedom to choose the operators $\hat{a}^{\dagger}_{j},\hat{a}_j$ to be bosonic or fermionic. In the following we assume that they are bosonic. 

	We are interested in the path integral representation of the Keldysh partition function of the problem \cite{kamenev_2011}. After analytically integrating out the bath degrees of freedom the expression has the following form:
\begin{align}
 Z & = \int D[\vect{a}] e^{ i \mathcal{S}^{}_{S} }_{} \prod\limits^{ \mathcal{N} }_{ n = 1 } e^{ i\mathcal{S}^{n}_{B,SB} }_{},
\end{align}
where $i\mathcal{S}_S$ origins from the system Hamiltonian $\hat{H}_S$ and $i\mathcal{S}^n_{B,SB}$ describes the effect of the $n$-th bath on the system. They are given by \cite{ishizaki_2009,tanimura_2012,Tanimura_1989_n, Tanimura_1990_n, Ishizaki_2005_n, Tanimura_2006_n}:
\begingroup
\allowdisplaybreaks
\begin{align}
 i \mathcal{S}^{}_S	= & \sum\limits^{ \mathcal{N} }_{n,n'=1} i \int^t_0 d\tau \vect{a}^{*T}_n (\tau) ( i \delta_{nn'} \partial_{\tau} - h_{nn'} ) \sigma_z \vect{a}_{n'} (\tau), \label{eq:S_S_action}\\
 \vect{a}_n (t) = & [ a^f_n(t) , a^b_n (t) ]^T,\\
 i \mathcal{S}^{n}_{B,SB}= & - \frac{1}{\pi} \int^{t}_{0} d\tau \int^{\tau}_{0} d\tau' n^{\times}_{n} (\tau) n^{\times}_{n} (\tau') \mathcal{F}_n(\tau - \tau') \nonumber  \label{eq:S_SB_action}\\
 & + \frac{i}{\pi} \int^{t}_{0} d\tau  \int^{\tau}_{0} d\tau' n^{\times}_{n}(\tau) n^{o}_{n} (\tau') \mathcal{D}_n (\tau - \tau'),\\
n^{\times}_{n} = & a^{f*}_{n}a^{f}_{n} - a^{b*}_{n} a^{b}_{n}, \label{eq:nX_Def} \\
n^{o}_{n} = & a^{f*}_{n}a^{f}_{n} + a^{b*}_{n} a^{b}_{n}, \label{eq:nO_Def}
\end{align}
\endgroup
where $a^{f}_{n}/a^{b}_n$ are the fields lying on the forward/backward part of the Keldysh contour and $\sigma_z$ is the Pauli $z$ matrix. The noise ($\mathcal{F}$) and dissipation ($\mathcal{D}$) kernels are given by:
\begingroup
\allowdisplaybreaks
\begin{align}
\mathcal{F}_n(t) 			& = \int d\omega J_n(\omega) \coth ( \omega/(2T) )  \cos (\omega t ),\\
\mathcal{D}_n(t) 			& = \int d\omega J_n(\omega) \sin (\omega t) ,\\
J_n(\omega)	& = \sum\limits_k \pi \lambda^2_{nk} \delta( \omega - \omega_{nk} ) ,
\end{align}
\endgroup
where $J_n (\omega)$ is the spectral density.

\subsection{Wigner formalism}
\label{subsec:Wigner_formalism}
We can express the expectation value of some system operator $\hat{O}$ in the following path integral form:
\begingroup
\allowdisplaybreaks
\begin{align}
{\rm tr}^{}_{S} \big[ \hat{O} \hat{\rho} (t) \big]
& = \int D[\vect{\psi}, \vect{\eta}] \mathcal{O}^{}_{ \mathcal{W} } (\vect{\psi}^{*}_t, \vect{\psi}^{}_{t}) e^{ i \mathcal{S} }_{}
				\mathcal{\rho}^{}_{ \mathcal{W} } ( \vect{\psi}^{*}_0 , \vect{\psi}^{}_{0} )  , \label{eq:Observable_exp_value_line_1} \\
D[\vect{\psi}, \vect{\eta}] & = \prod\limits^{}_{\tau} \prod\limits^{ \mathcal{N} }_{n=1} 
\frac{ d\Re \psi_{n,\tau} d\Im \psi_{n,\tau} d\Re \eta_{n,\tau} d\Im \eta_{n,\tau} }{\pi^2} , \label{eq:Observable_exp_value_line_2}
\end{align}
\endgroup
where $\psi_n/\eta_n$ are the quantum/classical fields which are obtained from the variable transformation 
\begin{align}
\label{eq:Wigner_trf}
a^{f/b}_{n} & = \psi_n \pm \frac{1}{2}\eta_n 
\end{align}
and $\mathcal{S} = \mathcal{S}^{}_{S} +  \sum^{}_{n} \mathcal{S}^{n}_{B,SB}$. The $i\mathcal{S}^n_{B,SB}$ term is defined in \eqref{eq:S_SB_action} where 
$ n^{\times}_{n} = \psi^*_{n} \eta_n + \eta^*_{n} \psi_n $, $ n^{o}_{n} 		= 2|\psi_n|^2_{} + \frac{1}{2}|\eta_n|^2_{} $ and $i\mathcal{S}_S$ is given by:
\begin{align}
i\mathcal{S}_S & = 
i\int d\tau [ \vect{\eta}^T (\tau) (	-i\partial_{\tau} - h^* ) \vect{\psi}^* (\tau) \nonumber \\
& \hspace{20.0mm} + \vect{\eta}^{*T} (\tau) ( i\partial_{\tau} - h ) \vect{\psi} (\tau)]
, \label{eq:action_sys_line_1} \\
\vect{f} (t) & = [f_{1} (t) \ldots f_{\mathcal{N}}(t) ]^T, \hspace{3.0mm} f \in \lbrace \psi,\eta \rbrace, \label{eq:action_sys_line_2}
\end{align}
where the elements $h_{nn'}$ of the matrix $h$ are given in \eqref{eq:Def_HS}. Eq. \eqref{eq:action_sys_line_1} can be obtained from $i\mathcal{S}_{S}$, defined in \eqref{eq:S_S_action}, by applying the variable transformation given in \eqref{eq:Wigner_trf} and then using the identity
\begin{align}
& \int^{t}_{0} d\tau \vect{\psi}^{*T} (\tau) \partial_{\tau} \vect{\eta} (\tau) = \nonumber \\
& \hspace{13.0mm} -\int^{t}_{0} d\tau \vect{\eta}^T (\tau) \partial_{\tau} \vect{\psi}^{*} (\tau) + \vect{\psi}^{*T} (\tau) \vect{\eta} (\tau) 
\Big\vert^{t}_{0},
\end{align} 
and neglecting the boundary terms. The Wigner distribution function $\rho^{}_{\mathcal{W}}$ and Weyl symbol of the operator $\mathcal{O}^{}_{\mathcal{W}}$ are defined as follows:
\begin{align}
& \mathcal{\rho}^{}_{ \mathcal{W} } ( \vect{\psi}^{*} , \vect{\psi} ) = 
	\prod\limits^{ \mathcal{N} }_{ n = 1} \int \frac{ d\Re \eta^{}_{n} d\Im \eta^{}_{n} }{ 2\pi^2 } \label{eq:rho_Wign} \\
& \times e^{ -\frac{1}{4}| \eta^{}_{n}|^{2}_{} - |\psi^{}_{n}|^{2}_{} + \frac{1}{2}( \psi^{*}_{n} \eta^{}_{n} - \eta^{*}_{n} \psi^{}_{n} ) }_{}
 \langle  \vect{\psi} + \frac{1}{2} \vect{\eta} | \hat{\rho}_S  | \vect{\psi} - \frac{1}{2} \vect{\eta} \rangle , \nonumber \\
&  \mathcal{O}^{}_{\mathcal{W}} (\vect{\psi}, \vect{\psi}^{*}_{}) = \prod \limits^{ \mathcal{N} }_{ n=1} \int \frac{ d\Re \eta^{}_{n} d\Im \eta^{}_{n} }{ 2 \pi  } \label{eq:O_Weyl} \\
&\hspace{29.0mm}  \times e^{ -\frac{1}{2}| \eta^{}_n |^2_{} }_{} \langle \vect{\psi} - \frac{1}{2}\vect{\eta} | \hat{O} | \vect{\psi} + \frac{1}{2} \vect{\eta}  \rangle, \nonumber
\end{align}
where the operator $\hat{O}$ is in its normal ordered form and $| \vect{\psi}\pm \frac{1}{2} \vect{\eta} \rangle $ 
is a coherent state with the property $ \hat{a}_n | \vect{\psi}\pm \frac{1}{2} \vect{\eta} \rangle =  ( \psi_n \pm \frac{1}{2} \eta_n ) | \vect{\psi}\pm \frac{1}{2} \vect{\eta} \rangle  $ and  $ \langle \vect{\psi} | \vect{\psi}'\rangle = \exp [ \sum^{\mathcal{N}}_{n=1} \psi^{*}_n  \psi'_n ] $. The same expression can be obtained if we represent the operator $\hat{O}$ in terms of symmetrized polynomials of $\hat{a}^{\dagger}_n$ and of $\hat{a}_n$ and then carry out the replacement $( \hat{a}^{\dagger}_n,\hat{a}_n ) \rightarrow ( \psi^*_n, \psi_n )$. The Wigner transform $\rho_{\mathcal{W}}$ of the density matrix is, in general, a pseudo-probability density function, i.e. it is normalized to one but it can take negative values. For the practical calculation of the path integral one can replace it with the following expression:
\begingroup
\allowdisplaybreaks
\begin{align}
\rho_{\mathcal{W}} (\vect{\psi}^*, \vect{\psi})	& =	N ( \rho_{\mathcal{W}} ) {\rm Sign } (\rho_{\mathcal{W}}, \vect{\psi}^*, \vect{\psi} )	\nonumber \\
	& \times \rho^{abs}_{\mathcal{W}} (\vect{\psi}^*, \vect{\psi}) , \\
\rho^{abs}_{\mathcal{W}} (\vect{\psi}^*, \vect{\psi})	& = |\rho_{\mathcal{W}} (\vect{\psi}^*, \vect{\psi})| / N ( \rho_{\mathcal{W}} ) , \label{eq:Def_rho_abs_W} \\
N ( \rho_{\mathcal{W}} ) & = \prod\limits^{ \mathcal{N} }_{n=1} \int d\Re \psi_n d\Im \psi_n | \rho_{\mathcal{W}}  (\vect{\psi}^*, \vect{\psi}) |, \\
{\rm Sign } (\rho_{\mathcal{W}}, \vect{\psi}^*, \vect{\psi} ) & = \left\lbrace
\begin{array}{rc}
1	& {\rm if }  \hspace{1.0mm} \rho^{}_{\mathcal{W}} (\vect{\psi}^*, \vect{\psi}) \geq	0 \\
-1	& {\rm if }  \hspace{1.0mm} \rho^{}_{\mathcal{W}} (\vect{\psi}^*, \vect{\psi}) <	0 \\
\end{array}
\right. .
\end{align}
\endgroup
In this case $\rho^{abs}_{\mathcal{W}}$ is a real probability density function. The derivation of Eqs. \eqref{eq:Observable_exp_value_line_1}, \eqref{eq:Observable_exp_value_line_2}, \eqref{eq:action_sys_line_1}, \eqref{eq:rho_Wign} and \eqref{eq:O_Weyl} is shown in \cite{Polkovnikov_2003}. 

	If we neglect the contribution $i\mathcal{S}_{B,SB}$ from the environment we can integrate out the $\eta$-variables by the use of the equation
\begin{align}
\label{eq:Eq_Delta_fct}
\int^{\infty}_{ - \infty } \frac{d\varphi}{\pi} e^{i 2 \varphi f}_{} & = \delta( f ), \hspace{3.0mm} f\in \mathbb{R}
\end{align}
for $\varphi \in \lbrace \Re\eta,\Im \eta \rbrace $. If we keep the $\psi_{n,0}$ variables fixed, the set of Dirac-delta function will define a unique path for the time evolution of the $\psi_{n,\tau}$ ($\tau>0$) variables. The path is described by the following equation:
\begin{align}
d\vect{\psi} (t) & = -ih \vect{\psi}(t)dt .
\end{align}
A single contribution to the observable ${\rm tr} [ \hat{O} \hat{\rho}(t) ]$ is given by:
\begin{align}
\label{eq:Single_contrib_to_observable}
\mathcal{O}_{\mathcal{W}} ( \vect{\psi}^*(t), \vect{\psi} (t) ) N(\rho_{\mathcal{W}}) {\rm Sign} ( \rho_{\mathcal{W}}, \vect{\psi}^*_0, \vect{\psi}_0 ).
\end{align}
To calculate the expectation value of $ {\rm tr} [ \hat{O} \hat{\rho}(t) ] $ we have to sample $ \psi_{n,0} $ from $\rho^{abs}_{\mathcal{W}}$ and calculate the mean value of \eqref{eq:Single_contrib_to_observable}. To take into account the effect of the environment, i.e. of $i\mathcal{S}_{B,SB}$, we have to apply to it additional transformations which will be explained in the next subsection.

\subsection{New mapping of the time non-local parts of the action to time local expressions}
\label{subsec:Mapping_proc}

	We will map the time non-local part of the action to a time local expression at the cost of introducing integrals over an additional set of variables which is a special realization of the idea proposed in \cite{stockburger2002exact}. To apply this mapping we assume that the noise ($\mathcal{F}$) and dissipation kernel ($\mathcal{D}$) of the action can always be represented as a sum of exponentially decaying functions:
\begin{align}
\frac{1}{\pi}\mathcal{D}(t) & = \sum\limits_{l\in L} 	\alpha^{D}_{l} e^{\lambda_l t},\\
\frac{1}{\pi}\mathcal{F}(t) & = \sum\limits_{l\in L}	\alpha^{F}_{l} e^{\lambda_l t}  
				+ \sum\limits_{l \in \tilde{L}} \alpha^{F}_{l} e^{ \lambda_l t} \hspace{10.0mm} \Re [ \lambda_l ] < 0,
\end{align}
where all $\alpha^D_l,\alpha^F_l$ constants are non-zero. It is important that the set of exponentially decaying functions in $\mathcal{D}$ is entirely included in the corresponding set of exponentially decaying functions in $\mathcal{F}$. This condition is automatically fulfilled if we assume that the spectral density is of the form $J(\omega) = \Theta(\omega) J'(\omega )$ with $J'(\omega)$ an odd function and with simple poles which do not lie on the real axis. Then by use of the residual theorem one can show that the set of poles in the upper half plane of $J'(\omega)$ is equal to $ \lbrace -i\lambda^{}_{l} \rbrace^{}_{l\in L} $ and the set of poles of $ \coth ( \omega/(2T) )$ is equal to $\lbrace -i\lambda_l \rbrace_{l \in \tilde{L} } $.


	The mapping of $i\mathcal{S}^n_{B,SB}$ given in \eqref{eq:S_SB_action} to a time-local action $i\mathcal{S}^{n,TL}_{B,SB}$ is defined as follows (we will denote the time arguments of the variables as subscripts):
\begin{align}
\exp \big[ iS^{n}_{B,SB} \big] & = 
\int D[\vect{x}^n] D[\vect{\phi}^n] D[\vect{\varphi}^n] \times  \nonumber \\ 
	& \times \exp \big[	iS^{n,TL}_{B,SB}	\big] \rho^n_{\Phi} (\vect{\phi}^n_0 ) , \label{eq:Wigner_identity_1_line_1} \\
iS^{n,TL}_{B,SB}	& = \sum\limits^{}_{\tau} i 2 \vect{\varphi}^{nT}_{\tau} \big( - \vect{\phi}^n_{\tau+\Delta t} + \vect{\phi}^n_{\tau}	+ {\scriptstyle \Delta}t A^n \vect{\phi}^n_{\tau} + \nonumber \\
+ {\scriptstyle \Delta}t \vect{v}^n n^{o}_{n,\tau} + &  \sqrt{ {\scriptstyle \Delta} t}  B^n\vect{x}^n_{\tau}	\big) 
+ i \sum\limits^{}_{\tau} {\scriptstyle \Delta}t  n^{\times}_{n,\tau} \vect{\varepsilon}^{nT}_{} \vect{\phi}^n_{\tau},\label{eq:Wigner_identity_1_line_2}\\
\int D[\vect{\varphi}^n]	& = \prod\limits^{}_{\tau} \prod\limits^{ \mathcal{M}_n }_{m=1} \int\limits^{\infty}_{-\infty} \frac{d\varphi^n_{m,\tau}}{\pi}, \\
\int D[\vect{\phi}^n]		& = \prod\limits^{}_{\tau} \prod\limits^{ \mathcal{M}_n }_{m=1} \int\limits^{\infty}_{-\infty} d\phi^n_{m,\tau},	\\
\int D[\vect{x}^n	]		& = \prod\limits^{}_{\tau} \prod\limits^{ \mathcal{M}_n }_{m=1} \int\limits^{\infty}_{-\infty} d x^n_{m,\tau} f_{X}(x^n_{m,\tau}), \label{eq:Dx_Def} \\
f_X(x) & =  \frac{1}{\sqrt{2\pi}} e^{-\frac{1}{2}x^2} , \label{eq:pdf_X}\\
B^n & = {\rm diag}[ b^{n}_1, \ldots , b^{n}_{\mathcal{M}_n} ], \hspace{3.0mm} b^{n}_{1}, \ldots b^n_{\mathcal{M}_n} \in \mathbb{R}_{\geq 0 }  \nonumber \\
A^n \in & \mathbb{R}^{ \mathcal{M}_n \times \mathcal{M}_n }_{}, \hspace{5.0mm} \vect{\varphi}^n_{\tau},  \vect{\phi}^n_{\tau}, \vect{v}^n, \vect{x}^n, \vect{\varepsilon}^n \in \mathbb{R}^{ \mathcal{M}_n }_{} 
, \nonumber 
\end{align}
The integer $\mathcal{M}_n$ is equal or larger than the number of elements in $ L \cup \tilde{L}$ and the $A^n$-matrix is chosen such that the set of its eigenvalues coincides with $ \lbrace \lambda^{}_l \rbrace^{}_{l \in L \cup \tilde{L}} $ (some eigenvalues may appear more than once). The initial distribution function of the $\phi$-fields is given by:
\begin{align}
\rho^n_{\Phi} (\vect{\phi}^n) & = \frac{ \exp [ -\frac{1}{2} \vect{\phi}^{nT}  (\Sigma^{n})^{-1} \vect{\phi}^n  ] }{ \sqrt{ (2\pi)^{\mathcal{M}_n} {\rm det}(\Sigma^n) } } , \label{eq:Initial_distribution}\\ 
\Sigma^n & = \int^{\infty}_{0} d\tau  \exp [ A^n\tau ] B^n B^{nT}  \exp [ A^{nT} \tau ]. \label{eq:Initial_distribution_Sigma}
\end{align}
The covariance matrix $\Sigma^n$ is well defined since all eigenvalues of $A$ have a negative real part. The exact values of $\vect{v}^n$, $\vect{\varepsilon}^n$, $B^n$, $A^n$ are determined after integrating out the $\vect{x}^n$, $\vect{\phi}^n$, $\vect{\varphi}^n$ variables in \eqref{eq:Wigner_identity_1_line_1} and comparing the result with \eqref{eq:S_SB_action}. Since the number of free variables is larger than the number of equations that have to be fulfilled, we have some freedom in the choice of $\vect{v}^n$, $\vect{\varepsilon}^n$, $B^n$, $A^n$. All details about the proof of \eqref{eq:Wigner_identity_1_line_1}, \eqref{eq:Wigner_identity_1_line_2} are given in the Appendix \ref{sec:App_Proof_1}, where we have assumed that the algebraric multiplicity of every eigenvalue of the $A^n$ matrix is equal to one. In this case the values of $\vect{v}^n$, $\vect{\varepsilon}^n$, $B^n$, $A^n$ are chosen such that the equation
\begin{equation}
\label{eq:Eq_coefficients}
\vect{\varepsilon}^{nT} e^{A^nt} \vect{v}^n =  \frac{1}{\pi} \mathcal{D}_n (t)
\end{equation}
is fulfilled. In addition, the constants have to be chosen such that the part of the expression
\begin{align}
m (t,t') & \overset{ t>t'}{=}  \int\limits^{ t' }_{0} d\tau \vect{\varepsilon}^{nT}_{} e^{ A^n( t - \tau) }_{} B^n B^{nT}_{} e^{ A^{nT}_{}(t' - \tau) }_{} \vect{\varepsilon}^n , \label{eq:Eq_coefficients_2}
\end{align}
that depends on the difference $(t-t')$ is equal to $\frac{1}{\pi} \mathcal{F}_n(t-t')$. In the following we will also add the constraint, that $v^{n}_m =0$ if $b^n_m=0$, which is not a necessary condition for \eqref{eq:Wigner_identity_1_line_1} to hold, but it will be needed to show \eqref{eq:Wigner_Identity_2a} and \eqref{eq:Wigner_Identity_2b}.

	The ansatz that we have used in \eqref{eq:Wigner_identity_1_line_1}, \eqref{eq:Wigner_identity_1_line_2} origins from the idea that different Hamiltonian operators of the environment can produce the same effective action $i\mathcal{S}_{B,SB}$ after integrating out the environmental degrees of freedom. One of the best known examples is the replacement of the Hamiltonian of each of the $\mathcal{N}$ environments in \eqref{eq:Def_H_BSB} with the Hamiltonian of a single harmonic oscillator (nuclear mode) that is coupled to one of the sites of the system and also to a Markovian bath, which is also composed of non-interacting harmonic oscillators \cite{Garg_1985}. If we integrate out the bath degrees of freedom from the path integral expression of the problem but keep the nuclear modes, then the contribution of the environment plus the nuclear mode will have the same form as the right-hand side of \eqref{eq:Wigner_identity_1_line_1}. The $\phi$ ($\varphi$) variables will correspond to the real or imaginary part of the classical (quantum) variables of the nuclear mode. Similarly to the $\eta$ variables in \eqref{eq:action_sys_line_1}, the $\varphi$ variables can be integrated out analytically by the use of \eqref{eq:Eq_Delta_fct} and the set of Dirac-delta functions will define a path for the time evolution of $\vect{\phi}^n_t$. The $x$-variables will be integrated by the use of Monte-Carlo methods, which effectively will make the equations for the $\phi$-variables stochastic. The noise in these equations origins from the Markovianity of the bath to which the nuclear mode was coupled. Since the dimension of the new $\vect{\varphi}^n_t$, $\vect{\phi}^n_t$, $\vect{x}^n_t$ in \eqref{eq:Wigner_identity_1_line_2} can be larger than two we can just assume that each site of the system is coupled to more than one nuclear modes, where the nuclear modes can be coupled to the same or to different Markovian environments.	
	
	Finally, we have to mention that $ \rho^n_{\Phi}(\vect{\phi})$ can be interpreted as the probability density function, where all nuclear modes are in equilibrium with the rest of the environment for the case that they are not coupled to the system. To demonstrate this we can decouple the nuclear modes from the system in the interval $[-t_R,0]$ ($-t_R<0$) by setting $\vect{\varepsilon}^n=\vect{v}^n=0$ in \eqref{eq:Wigner_identity_1_line_2} and let them evolve until they reach a steady state at $t=0$. The equation of motion for the $\phi^n_{m,\tau}$ variables which is derived after integrating out the $\varphi^n_{m,\tau}$ variables by using \eqref{eq:Eq_Delta_fct}, is given by:
\begin{align}
\vect{\phi}^{n}_{t+\Delta t}	& = \vect{\phi}^{n}_{t} + {\scriptstyle \Delta}t A^n \vect{\phi}^n_t + \sqrt{{\scriptstyle \Delta}t} B \vect{x}^n_t.
\end{align}
The solution of this equation at $t=0$ is given by:
\begin{align}
\vect{\phi}^n_{0} & =	
e^{A^n t_R} \vect{\phi}^n_{-t_R} 
+ \sum\limits^{}_{0 \geq \tau \geq -t_R} e^{-A^n \tau} B \sqrt{ {\scriptstyle \Delta}t } \vect{x}^n_{\tau} \nonumber \\
& \rightarrow \sum\limits^{}_{0 \geq \tau} e^{-A^n \tau} B \sqrt{ {\scriptstyle \Delta}t } \vect{x}^n_{\tau},
\end{align}
where in the second line we have taken the limit $-t_R \rightarrow -\infty$. In the first (second) line the sums over $\tau$ is from $-t_R$ ($-\infty $) to $0$ by taking steps of length ${\scriptstyle \Delta}t$. By interpreting every $x^n_{m,t}$ as a realization of a normally distributed random variable $X^n_{m,t} \sim \mathcal{N}(0,1)$ it follows that the steady state solution is also a random variable with zero mean and variance equal to
\begin{align}
\langle \vect{\phi}^n_0 \vect{\phi}^{n*T}_0 \rangle	& = \sum\limits^{}_{0 \geq \tau} e^{-A^n \tau} B^n B^{nT} e^{-A^{nT} \tau} {\scriptstyle \Delta}t,
\end{align}
where we have used that $\langle \vect{X}^n_{t} \vect{X}^{nT}_{t'} \rangle  = \delta_{t,t'} \mathbbm{1}$. The last expression is equal to the definition of the covariance matrix of the multivariate probability density function $\rho^n_{\Phi}$ in \eqref{eq:Initial_distribution}, which proves our statement.

\subsection{Truncated Wigner approximation}
\label{subsec:TWA_new}

	We will conclude this section by giving the equations of motion for the $\psi_{n,\tau}$, $\phi^n_{m,\tau}$-variables, which are obtained by applying the TWA. For the current problem this approximation is equivalent to neglecting the $\frac{1}{2} |\eta_n|^2 $-terms in $n^{o}_{n}$, defined in \eqref{eq:nO_Def}, which appear in every $\mathcal{S}^{n,TL}_{B,SB}$ contribution to the action. We can integrate out the $\eta_n$, $\varphi^n_{m,\tau}$ terms by the use of \eqref{eq:Eq_Delta_fct}. If we keep the variables $x^{n}_{m,\tau}$, $\phi^n_{m,0}$, $\psi_{n,0}$ fixed, the set of Dirac-delta functions defines a unique path for the time evolution of the $\psi_{n,\tau}$  and $\phi^n_{m,\tau}$ variables, which is given by:
\begingroup
\allowdisplaybreaks
\begin{align}
\vect{\psi}_{t+\Delta t}		& = \vect{\psi}_t	- i \tilde{h}(t) \vect{\psi}_t {\scriptstyle \Delta } t, \label{eq:Path_psi_TWA} \\
\vect{\phi}^{n}_{t+\Delta t}	& = \vect{\phi}^n_t	+ A^{n}_{} \vect{\phi}^n_t {\scriptstyle \Delta } t	+ 2 |\psi_{n,t}|^2 \vect{v}^n {\scriptstyle \Delta} t \nonumber \\
								& +  B^n \vect{x}^n_t \sqrt{ {\scriptstyle \Delta} t }, \label{eq:Path_phi_TWA}\\
\tilde{h}( t)	& = h - {\rm diag} \big[ \vect{\varepsilon}^{1T} \vect{\phi}^{1}_{t}, \ldots \vect{\varepsilon}^{ \mathcal{N} T} \vect{\phi}^{ \mathcal{N} }_{t}  \big]. \label{eq:Langevin_Semiclass_3}
\end{align}
\endgroup
This is the Euler-Maruyama discretization of the following set of stochastic equations:
\begingroup
\allowdisplaybreaks
\begin{align}
d\vect{\psi}(t)			& = -i\tilde{h}(t) \vect{\psi}(t) dt,	\label{eq:Langevin_Semiclass_1} \\
d\vect{\phi}^{n}_{}(t)	& = A^{n}_{} \vect{\phi}^{n}_{}(t) dt +  \vect{v}^{n}_{} 2 |\psi_n (t)|^2 dt 	\nonumber \\
						& + B^{n}_{} d\vect{W}^{n}_{}(t), \label{eq:Langevin_Semiclass_2} \\
\vect{W}^{n}_{} (t)	& = [ W^n_1 (t), \ldots W^n_{\mathcal{M}_n} (t) ]^T ,	\label{eq:Def_Wiener_vect}
\end{align}
\endgroup
where the set $\lbrace W^n_m (t) \rbrace $ ($n\in \lbrace 1 \ldots \mathcal{N} \rbrace $, $m \in \lbrace 1 \ldots \mathcal{M}_n \rbrace $) is a set of independent Wiener processes, which have the property $dW^n_m (t) dW^{n'}_{m'} (t) = dt \delta_{nn'} \delta_{mm'}$. To calculate the expectation value of $ {\rm tr} [ \hat{O} \hat{\rho}(t) ] $ we have to sample the variables $ x^{n}_{m,\tau}$ from a normal distribution $\mathcal{N}(0,1)$ and the variables $ \phi^n_{m,0}$, $\psi_{n,0} $ from $\rho^n_{\Phi}$, $\rho^{abs}_{\mathcal{W}}$, defined in \eqref{eq:Initial_distribution}, \eqref{eq:Def_rho_abs_W}, and calculate the mean value of \eqref{eq:Single_contrib_to_observable}.

\section{Stochastic equation with all quantum corrections}
\label{sec:Stochastic_equation_with_all_quantum_corrections}
In this section we will analytically integrate all $ \varphi $- and $ \eta $-variables from \eqref{eq:Observable_exp_value_line_1} without neglecting any terms in the action.  The idea is to make all terms in $i \mathcal{S}^{n}_{B,SB}$ ($n \in \lbrace 1\ldots \mathcal{N} \rbrace $) to be linear in $(\varphi,\Re \eta , \Im \eta)$ and also completely imaginary, which will allow us to integrate them out via \eqref{eq:Eq_Delta_fct}. To do this we will introduce first a set of probability density functions $f_{R|X}(r|x),f_{\Theta} (\theta)$: 
\begingroup
\allowdisplaybreaks
\begin{align}
f^{}_{\Theta} (\theta)	& = \left\lbrace
\begin{array}{cc}
1/(2\pi)	& {\rm if} \hspace{1.0mm} \theta \in	[0,2\pi] \\
0			& {\rm if} \hspace{1.0mm} \theta \notin	[0,2\pi]
\end{array}
\right. , \label{eq:pdf_Theta} \\
f^{}_{R|X} (r|x)	& = \left\lbrace
\begin{array}{ll}
\int^{\infty}_{0} d\rho \rho r e^{-\frac{1}{2}(\rho^4 + 2\rho^2x)}_{} J_0 (\rho r)	& {\rm if} \hspace{1.0mm} r \geq 0 \\
0	& {\rm if} \hspace{1.0mm} r<0
\end{array}
\right. , \label{eq:pdf_RX} \\
J_{0}(x)	& = \int^{2\pi}_{0}  \frac{d\varepsilon}{2\pi} \exp [ -i x \sin (\varepsilon) ],
\end{align}
\endgroup
where $J_0 (x)$ is the Bessel function of first kind. They are defined such that the following equation is true for every $n \in \lbrace 1 , \ldots \mathcal{N} \rbrace $:
\begin{widetext}
\begin{align}
& \int
D[\vect{x}^n] 
D[\vect{\phi}^n] 
D[\vect{\varphi}^n] 
\exp \Big[ i\mathcal{S}^{n,TL}_{B,SB}	\Big] = 
\int 
D[\vect{x}^n] 
D[\vect{\phi}^n] 
D[\vect{\varphi}^n] 
D[\vect{\theta}^n]
D[\vect{r}^n]
\exp \Big[	i\tilde{\mathcal{S}}^{n,TL}_{B,SB} \Big], \label{eq:Wigner_Identity_2a} \\
& i\tilde{\mathcal{S}}^{n,TL}_{B,SB}	= \sum\limits^{}_{\tau} i2\vect{\varphi}^{nT}_{\tau}\big(	
-\vect{\phi}^n_{\tau + \Delta t} + \vect{\phi}^n_{\tau}	 + {\scriptstyle \Delta} t A^n \vect{\phi}^n_{\tau} + {\scriptstyle \Delta} t \vect{v}^n 2 |\psi_{n,\tau}|^2 + \sqrt{ {\scriptstyle \Delta} t} \tilde{B}^n \vect{x}^n_{\tau}
\big) + \sum\limits^{}_{\tau} i {\scriptstyle \Delta} t n^{\times}_{n,\tau} \vect{\varepsilon}^{nT} \vect{\phi}^n_{\tau} \nonumber \\
& \hspace{12.0mm} 
+ \sum\limits^{}_{\tau}	 i 2 \Re \Big(		\eta^*_{n,\tau} \sum\limits^{\mathcal{M}_n}_{ \substack{ m=1\\ b^n_m>0 } } \Big( \frac{|v^n_m|}{ 2b^n_m } \Big)^{1/2} {\scriptstyle \Delta } \chi^n_{m,\tau}	\Big). \label{eq:Wigner_Identity_2b}
\end{align}
\end{widetext}
The sum in the second line of \eqref{eq:Wigner_Identity_2b} is taken only over those $m \in \lbrace 1, \ldots \mathcal{M}_n \rbrace $, where $b^n_m>0$. The new differentials and variables are defined as follows:
\begingroup
\allowdisplaybreaks
\begin{align}
D[\vect{\theta}^n]	& = \prod\limits^{}_{\tau} \prod\limits^{ \mathcal{M}_n }_{m=1} d\theta^n_{m,\tau} f^{}_{\Theta} (\theta^n_{m,\tau}) , \\
D[\vect{r}^n]		& = \prod\limits^{}_{\tau} \prod\limits^{ \mathcal{M}_n }_{m=1} dr^n_{m,\tau} f^{}_{R|X} (r^n_{m,\tau}|x^n_{m,\tau}), \\
{\scriptstyle \Delta } \chi^n_{m,t}	& = ({\scriptstyle \Delta } t)^{1/4} r^n_{m,t} e^{i \theta^n_{m,t}}_{}/2, \\
\tilde{B}^n		& = {\rm diag}[ -{\rm Sign}(v^n_1)b^n_1, \ldots,  -{\rm Sign}(v^n_{\mathcal{M}_n})b^n_{\mathcal{M}_n} ].
\end{align}
\endgroup
The proof of \eqref{eq:Wigner_Identity_2a} is given in the Appendix \ref{sec:App_proof_linearize_action_2}. Here we will only point out that the right-hand side of \eqref{eq:Wigner_Identity_2a} differs from the TWA, applied to the left-hand side of the same equation, only by the ${\scriptstyle \Delta} \chi$-term in $i\tilde{S}^{n,TL}_{B,SB}$. If we neglect the ${\scriptstyle \Delta} \chi$ term in the second line of \eqref{eq:Wigner_Identity_2b}, then we can integrate out the $\theta^n_{m,t}$, $r^n_{m,t}$ variables in the path integral on the right-hand side of \eqref{eq:Wigner_Identity_2a} since nothing will depend on them. In addition, via the transformation $x^{n}_{m,\tau} \rightarrow -{\rm Sign}(v^n_m) x^{n}_{m,\tau} $ at the right-hand side of \eqref{eq:Wigner_Identity_2a}, the $D[\vect{x}^n]$ term does not change, but the $ \sqrt{{\scriptstyle \Delta }t} \tilde{B}^n \vect{x}^n_{\tau}$ term in \eqref{eq:Wigner_Identity_2b} transforms to $\sqrt{{\scriptstyle \Delta }t} B^n \vect{x}^n_{\tau}$ which proves our statement.

\begin{figure}
	\centering
	\includegraphics[width=0.9\linewidth]{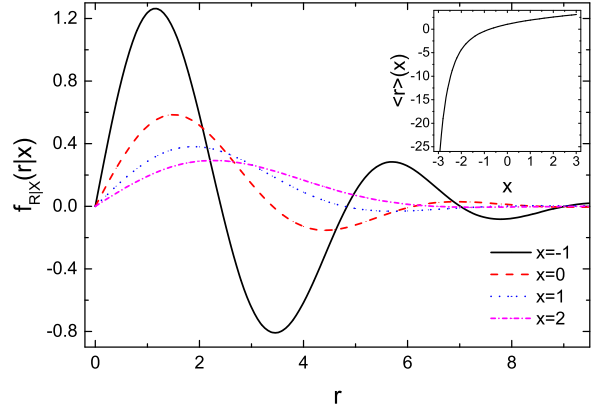}
	\caption{ (Color online) Pseudo-probability density function $f_{R|X}(r|x)$ for x=-1 (black solid), x=0 (red dashed), x=-1 (blue dotted) and x=2 (magenta dash dotted). Inset: first moment of the probability density function.}
	\label{fig:Graph_Pdf}
\end{figure}

	We have to remind, that $f_{R|X}$ is a pseudo-probability density function, as shown in Fig. \ref{fig:Graph_Pdf}, and it can be replaced by the following expression:
\begingroup
\allowdisplaybreaks
\begin{align}
f_{R|X} 		(r|x) 		& = N(f_{R|X},x) {\rm Sign}(f_{R|X},r,x) \nonumber \\
							& \times f^{abs}_{R|X} (r|x) , \\
f^{abs}_{R|X}	(r|x)		& = |f_{R|X} (r|x)|/N(f_{R|X},x),	\label{eq:pdf_f_abs_RX} \\
N(f_{R|X},x)				& = \int dr | f_{R|X} (r|x) |, \\
{\rm Sign} (f_{R|X},r,x)	& = \left\lbrace
\begin{array}{rl}
1	& {\rm if} \hspace{1.0mm} f_{R|X}(r|x) \geq 0 \\
-1	& {\rm if} \hspace{1.0mm} f_{R|X}(r|x) < 0
\end{array}
\right. .
\end{align}
\endgroup

	With the formula \eqref{eq:Eq_Delta_fct} we can integrate out all $\varphi^{n}_{m,\tau}$, $\Re \eta_{n,\tau}$, $\Im \eta_{n,\tau}$  variables from the path integral representation of the expectation value of the operator $\hat{O}$. If we keep the variables $x^n_{m,\tau}$, $r^n_{m,\tau}$, $\theta^n_{m,\tau}$, $\phi^n_{m,0}$, $\psi_{n,0}$ fixed, the set of Dirac-delta functions defines the following equation for the time evolution of $\vect{\psi}_{\tau}$, $\vect{\phi}^n_{\tau}$:
\begin{align}
\vect{\psi}_{t+\Delta t} & = \vect{\psi}_t - i \tilde{h}(t) \vect{\psi}_t {\scriptstyle \Delta} t
+ \sum\limits^{ \mathcal{N} }_{n=1} \sum\limits^{ \mathcal{M}_n }_{m=1} \vect{\kappa}^{nm}_{} {\scriptstyle \Delta} \chi^{n}_{m,t}, \label{eq:Path_psi_exact} \\
\vect{\phi}^{n}_{t+\Delta t} &  = \vect{\phi}^n_t 
								+ A^n \vect{\phi}^n_t {\scriptstyle \Delta} t 
								+ 2 |\psi_{n,t}|^2 \vect{v}^n {\scriptstyle \Delta} t \nonumber \\
							 &	+ \tilde{B}^n \vect{x}^n_t \sqrt{{\scriptstyle \Delta} t}, \label{eq:Path_phi_exact} \\
\vect{\kappa}^{nm}_{}	& = \Big[ 0 \ldots  \Big( \frac{|v^n_m|}{2b^n_m} \Big)^{1/2} \ldots 0 \Big]^T \in \mathbb{R}^{\mathcal{N}} ,
\end{align}
where only the $n$-th element of the $\vect{\kappa}^{nm}$ vector is non-zero. The last sum in \eqref{eq:Path_psi_exact} is taken only over those $(n,m)$, where $b^n_m>0$. The equations \eqref{eq:Path_psi_TWA}, \eqref{eq:Path_phi_TWA}, obtained from the TWA, differ from \eqref{eq:Path_psi_exact}, \eqref{eq:Path_phi_exact} only by the $B^n$ matrix, which is replaced by $\tilde{B}^n$ and by the absence of the term proportional to ${\scriptstyle \Delta} \chi $. To calculate the expectation value of the operator $\hat{O}$ we sample the variables $ x^{n}_{m,\tau}$, $\theta^{n}_{m,\tau}$, $r^{n}_{m,\tau}$, $\phi^n_{m,0}$, $\psi^{}_{n,0}$ from the probability density functions $f_X$, $f_{\Theta}$, $f^{abs}_{R|X}$, $\rho^n_{\Phi}$, $\rho^{abs}_{\mathcal{W}}$ given in \eqref{eq:pdf_RX},
\eqref{eq:pdf_Theta},
\eqref{eq:pdf_f_abs_RX},
\eqref{eq:Initial_distribution},
\eqref{eq:Def_rho_abs_W} and calculate the mean value of:
\begin{align}
& \mathcal{O}_{\mathcal{W}} (\vect{\psi}^*_0,\vect{\psi}_0) N(\rho_{\mathcal{W}}) {\rm Sign} (\rho_{\mathcal{W}}, \vect{\psi}^*,\vect{\psi} ) \times  \\
& \times \prod\limits^{}_{\tau} \prod\limits^{ \mathcal{N} }_{n=1} \prod\limits^{ \mathcal{M}_n }_{m=1} N ( f_{R|X}, x^{n}_{m,\tau} ) {\rm Sign} (f_{R|X},r^{n}_{m,\tau},x^{n}_{m,\tau}). \nonumber
\end{align}
Since all normalisation constants $ N ( f_{R|X}, x^{n}_{m,\tau} ) $ are always larger than one, it follows that their product will grow exponentially in time. This growth has to be compensated by the exponential decay of $ \mathcal{O}^{}_{\mathcal{W}} $ and/or by the alternating sign of the product of all sign-functions of every sampling of the random variables. In both cases this will require an exponential growth of the number of trajectories to obtain a good approximation of the expectation value. An additional problem comes from the ${\scriptstyle \Delta} \chi $-term in \eqref{eq:Path_psi_exact}, because it is proportional to $({\scriptstyle \Delta}t)^{1/4}$ and this requires the use of very small time steps to obtain an accurate trajectory. Although being practically inapplicable, this new representation of the problem is a good starting point for the derivation of corrections that go beyond the truncated Wigner approximation.

\section{Corrections of the TWA}
\label{sec:TWAQC}
	
	To obtain an approximation that can be applied to the calculation of practical problems, we have to find a way to eliminate the pseudo-probability density function $f_{R|X}$ from the path integral representation of ${\rm tr}[\hat{O} \hat{\rho}(t)]$. To do this we take into account the fact, that we are only interested in the time evolution of the different elements of the density matrix. Since the system lies in the single exciton subspace, the density matrix and its elements are given by:
\begingroup
\allowdisplaybreaks
\begin{align}
\hat{\rho}_S	& = \sum_{nn'} \rho_{nn'} \hat{a}^{\dagger}_n |0\rangle \langle 0| \hat{a}_{n'},	\label{eq:Def_rho_1ex_space}	\\
\rho_{nn'}		& = {\rm tr}_S[\hat{a}^{\dagger}_{n'} \hat{a}^{}_{n} \hat{\rho}_S (t) ].			\label{eq:Def_rho_el_1ex_space}
\end{align}
\endgroup
The Weyl symbol of the $\hat{a}^{\dagger}_{n'} \hat{a}^{}_{n}$ operator is equal to:
\begin{align}
\mathcal{O}_{\mathcal{W}} (t) & = \psi_{n}(t) \psi^*_{n'}(t) - \delta_{nn'}/2 .
\end{align}
An equation of motion for $\psi_n(t) \psi^*_{n'}(t)$ can be obtained if we take the equation for $\vect{\psi}$ from \eqref{eq:Path_psi_exact} and derive from it the equation for $\vect{\psi}^{*T}$. Then we take the product of both equations such that on the left-hand side of the equation we obtain the matrix $ \vect{\psi} \vect{\psi}^{*T}$. In the new equation we replace all polynomials in ${\scriptstyle \Delta} \chi$ with their average over $\theta$ and $r$. Up to the fourth moment the only non-vanishing expectation values are
\begingroup
\allowdisplaybreaks
\begin{align}
\int dr d\theta f_{R|X} (r|x) f_{\Theta} (\theta) {\scriptstyle \Delta} \chi {\scriptstyle \Delta} \chi^*		& = \sqrt{ {\scriptstyle \Delta}t } x, \label{eq:Moment_A} \\
\int dr d\theta f_{R|X} (r|x) f_{\Theta} (\theta) ( {\scriptstyle \Delta} \chi {\scriptstyle \Delta} \chi^* )^2	& = 2 {\scriptstyle \Delta} t( x^2 -1 ).
\end{align}
\endgroup
The last result can be explained by the properties of the $f^{}_{\Theta}$-function and the phase $\exp [ i \theta^{n}_{m,\tau} ]$, which leaves only those expectation values to be non-zero, where every ${\scriptstyle \Delta} \chi$ is multiplied with its complex conjugate. In this case the expression does not depend on the $\theta$-variable. After taking the average over all $r^{n}_{m,\tau}$, $\theta^n_{m,\tau}$ variables, the equation of motion for the matrix $\vect{\psi} \vect{\psi}^{*T}$ has the following form:
\begin{align}
\vect{\psi}^{}_{t+\Delta t} \vect{\psi}^{*T}_{t+\Delta t} & = (1-i {\scriptstyle \Delta} t \tilde{h}(t)) \vect{\psi}^{}_{t} \vect{\psi}^{*T}_{t} (1+i\Delta t \tilde{h}(t)) \nonumber \\
	& + \sum\limits^{\mathcal{N}}_{n=1} \sum\limits^{ \mathcal{M}_n }_{m=1} \vect{\kappa}^{nm}_{} \vect{\kappa}^{nm *T}_{} x^{n}_{m,t} \sqrt{ {\scriptstyle \Delta}t } .
\end{align}
If we take the limit ${\scriptstyle \Delta}t \rightarrow 0$ and neglect the terms proportional to ${\scriptstyle \Delta}t^s$ ($s>1$), the equation is equal to the Euler-Maruyama discretization of the following stochastic equation:
\begin{align}
d(\vect{\psi} \vect{\psi}^{*T}) (t)
						& = [-i\tilde{h}(t),(\vect{\psi} \vect{\psi}^{*T}) (t)]dt \nonumber \\
						& + \sum\limits^{\mathcal{N}}_{n=1} \sum\limits^{\mathcal{M}_n}_{m=1} \vect{\kappa}^{nm}_{} \vect{\kappa}^{nm *T}_{} dW^{n}_{m}(t) , \label{eq:psi_psi_dag_continuous} \\
d\vect{\phi}^n (t)		& = A^n \vect{\phi}^n(t) dt + \vect{v}^n 2 ( \vect{\psi} \vect{\psi}^{*T} )_{nn}(t) dt \nonumber \\
						& + \tilde{B}^n d\vect{W}^n(t) , \label{eq:phi_continuous}
\end{align}
where $\vect{W}^n (t)$ is defined in \eqref{eq:Def_Wiener_vect}. A single realization of the observable is obtained by sampling $\phi^n_{m,0}$, $\psi_{n,0}$ from $\rho^n_{\Phi}$, $\rho^{abs}_{\mathcal{W}}$ and calculating the mean value of \eqref{eq:Single_contrib_to_observable}. An alternative way to derive \eqref{eq:psi_psi_dag_continuous} and the possibility to calculate multitime correlation functions within this approximation is given in the Appendix \ref{sec:App_alt_deriv_CTWA}.
	
	If we neglect the second line of \eqref{eq:psi_psi_dag_continuous}, we obtain again the TWA, since $(\vect{\psi} \vect{\psi}^{*T})(t)$ has a solution of the form
\begin{align}
(\vect{\psi}\vect{\psi}^{*T})(t) & = 
\Big(	T e^{-i\int^{t}_{0} \tilde{h}(\tau) d\tau }	\Big) (\vect{\psi}\vect{\psi}^{*T})(0)
\Big(	T^{\dagger} e^{i\int^{t}_{0} \tilde{h}(\tau) d\tau }	\Big),
\end{align}
which factorizes into a product of the solutions for $\vect{\psi}$ and $\vect{\psi}^*$ from \eqref{eq:Langevin_Semiclass_1}.
	
	If we use the solution of \eqref{eq:phi_continuous}, which is equal to \eqref{eq:phi_path_solution} after replacing $n^o(\tau)$ with $2 ( \vect{\psi} \vect{\psi}^{*T} )_{nn} (\tau)$, in the definition of $\tilde{h}(t) = h - {\rm diag}[\vect{\varepsilon}^{1T}\vect{\phi}^1(t), \ldots  \vect{\varepsilon}^{\mathcal{N} T}\vect{\phi}^{ \mathcal{N} }(t) ]$, then \eqref{eq:psi_psi_dag_continuous} transforms into a differential equation with a memory kernel. This memory kernel plays a crucial role for the effect of the new ${\scriptstyle \Delta}\chi$-dependent term on the evolution of the system. If the memory term is absent, then the noise generated from the ${\scriptstyle \Delta}\chi$-terms will not have any effect on the $\vect{\psi}\vect{\psi}^{*T}$ matrix on average.

	In the end we mention that \eqref{eq:psi_psi_dag_continuous} preserves the trace of the $ \vect{\psi} \vect{\psi}^{*T}_{} $-matrix on average (and also the trace of $\rho^{}_{}  = \langle \vect{\psi} \vect{\psi}^{\dagger}_{} \rangle - \mathbbm{1}/2 $). Since $ \vect{\kappa}^{nm}_{} \vect{\kappa}^{nm\dagger}_{}$ is a diagonal matrix one can show by induction that 
\begin{align}
{\rm tr} \big[ \big( \vect{\psi} \vect{\psi}^{*T} \big) (t) \big] & = {\rm tr} \big[ \vect{\psi}_0 \vect{\psi}^{*T}_0 \big] 
+ \nonumber \\
	& + \sum\limits^{ \mathcal{N} }_{n=1} \sum\limits^{ \mathcal{M}_n }_{m=1} \Big( \frac{|v^n_m|}{2b^n_m} \Big)^2  \sum\limits^{}_{\tau} \sqrt{ {\scriptstyle \Delta} t } x^n_{m,\tau},
\end{align}
which is zero on average.

\section{Examples}
\label{sec:Examples}
We consider the example of having a system of $\mathcal{N} = 2$ sites, each of them being linearly coupled to an independent bath of harmonic oscillators. The corresponding spectral densities will be the same and are given by:
\begingroup
\allowdisplaybreaks
\begin{align}
J(\omega)	& = J'(\omega) \Theta(\omega), \\
J'(\omega)	& = \frac{ 2a^{}_{1} \omega^{3}_{} + 2\big[ a^{}_1(\gamma^2_{} \mp \Omega^2_{} ) + 2a^{}_2 \gamma \Omega  \big]\omega }{ 
[ \gamma^{2}_{} \pm \Omega^{2}_{} - \omega^2_{} ]^{2}_{} + 4\gamma^2_{} \omega^{2}_{}	} ,
\end{align}
\endgroup
where $\Theta $ is the Heaviside step function. Since $J'(\omega)$ is odd, we can use the residual theorem to calculate the noise and dissipation kernels  of the action $i\mathcal{S}^n_{B,SB}$ $(n=1,2)$:
\begingroup
\allowdisplaybreaks
\begin{align}
\frac{1}{\pi} \mathcal{D}(t) & = \big[ a^{}_{1}C(\Omega t) + a^{}_2 S(\Omega t) \big] e^{-\gamma t}_{} ,  \label{eq:Diss_part_example} \\
\frac{1}{\pi} \mathcal{F}(t) & = 2T ( a^{}_1 f^{}_S +   a^{}_2 f^{}_{A}  )C(\Omega t) e^{-\gamma t}_{} \nonumber \\
               & + 2T ( a^{}_2 f^{}_S \mp a^{}_1 f^{}_{A}  )S(\Omega t) e^{-\gamma t}_{} \nonumber \\
 			   & + \sum\limits^{ \infty }_{ l=1 } 2Ti J'( i\nu^{}_{l} ) e^{-\nu^{}_{l} t }_{} , \label{eq:Def_F_wieder} \\
( C,S ) & = \left\lbrace
\begin{array}{l}
(	\cos  	,	\sin			) 		\\
(	\cosh 	,	\sinh			)
\end{array}\right. 
, \\
f^{}_{S} & = \gamma \bigg[ \frac{1}{ \gamma^2_{} \pm \Omega^{2}_{} }  + \sum\limits^{ \infty}_{l=1} 2 \frac{ \gamma^{2}_{} \pm \Omega^2_{} - \nu^{2}_{l} }{ [ \gamma^2_{} \pm \Omega^2_{} - \nu^2_l ]^2_{} \pm 4\nu^{2}_{l}\Omega^2_{} } \bigg], \label{eq:Def_fS}\\
f^{}_{A} & = \Omega \bigg[ \frac{1}{ \gamma^2_{} \pm \Omega^{2}_{} }  + \sum\limits^{ \infty}_{l=1} 2 \frac{ \gamma^{2}_{} \pm \Omega^2_{} + \nu^{2}_{l} }{ [ \gamma^2_{} \pm \Omega^2_{} - \nu^2_l ]^2_{} \pm 4\nu^{2}_{l}\Omega^2_{} } \bigg], \label{eq:Def_fA} \\ 
 \nu^{}_{l} & = 2\pi l T, \hspace{4.0mm} \Omega,\gamma, a^{}_1, a^{}_{2} \in \mathbb{R}, \hspace{4.0mm} \gamma,\Omega >0. \nonumber
\end{align}
\endgroup
The $(C(\Omega t),S(\Omega t))$ functions are replaced with their upper/lower definition, when we use the upper/lower sign of $\pm,\mp$ in the definition of $J'(\omega)$. If we use the lower definition of $(C(t),S(t))$, we have the additional constraint, that $\gamma>\Omega$. Depending on the situation we will use the $a'_1$, $a'_2$ instead of $a_1$, $a_2$. Both pairs of variables are given by:
\begin{equation}
\label{eq:New_coupl_const}
a^{}_{1} = \frac{\gamma^2_{} \pm \Omega^2_{} }{ \gamma } a'_{1}, \hspace{5.0mm}
a^{}_{2} = \frac{\gamma^2_{} \pm \Omega^2_{} }{ \Omega } a'_{2}
\end{equation}
and are related to the reorganization energy as follows:
\begin{equation}
\sum_k  \frac{\lambda^2_{nk}}{ \omega_{nk} } =  a'_1 + a'_2,  \hspace{8.0mm} n \in \lbrace 1 ,\ldots \mathcal{N} \rbrace.
\end{equation}
For simplicity we approximate ${\coth}(\frac{\omega}{2T}) \approx \frac{2T}{\omega}$ and neglect all sums over the $l$-index in \eqref{eq:Def_F_wieder}, \eqref{eq:Def_fS}, \eqref{eq:Def_fA}, which is justified at high temperatures. In this case both functions $\mathcal{F}$ and $\mathcal{D}$ can be decomposed as a sum of two exponentially decaying functions with the exponents
\begin{align}
\lambda_{1,2} & = 
\left\lbrace
\begin{array}{ll}
-\gamma \pm i \Omega & {\rm for} \hspace{1.0mm}	(C,S)=(\cos , \sin ) \\
-\gamma \pm  \Omega & {\rm for} \hspace{1.0mm}	(C,S)=(\cosh ,\sinh )
\end{array}
\right. . \label{eq:Eigenvalues}
\end{align}
The time non-local part of the action $i\mathcal{S}^n_{B,SB}$, generated from every one of the reservoirs, can be mapped to a time local action in the same way as described in \ref{subsec:Mapping_proc}. Additional information about the choice of $A^n, \vect{\varepsilon}^n, \vect{v}^n, B^n$ used in \eqref{eq:Wigner_identity_1_line_1},\eqref{eq:Wigner_identity_1_line_2} can be found in the Appendix \ref{sec:App_details_DA_model}.

	We will assume that both spectral densities are equal and we will consider the two cases, where $a'_1=0$ or $\Omega = 0$. Both spectral densities now have the following form:
\begin{align}
\label{eq:Spec_fct_both}
J'(\omega) & = 
\left\lbrace
\begin{array}{ll}
a'^{}_2 \frac{ {\textstyle 4( \gamma^2_{} \pm \Omega^2_{} )\gamma \omega } }{ {\textstyle ( \gamma^2_{} \pm \Omega^2_{} - \omega^2_{} )^2_{}  + 4\gamma^2_{}\omega^2_{} } }
& a'^{}_1 =0
\\
a'^{}_1 \frac{ {\textstyle 2\gamma\omega } }{ {\textstyle \gamma^2_{} + \omega^2_{} } } & \Omega = 0
\end{array}
\right. .
\end{align}

	All parameters of the system will be given in units of the difference between the energy levels of the donor and the acceptor $\Delta = h_{11}-h_{22}$. The coupling between both sites is set to $h_{12}/\Delta = 0.4$ and the temperature of the reservoirs is set to $T/\Delta = 2$, unless it is not mentioned explicitly. We assume, that the excitation is initially localized at the first site, which produces the following Wigner distribution function:
\begingroup
\allowdisplaybreaks
\begin{align}
\rho^{}_{\mathcal{W}} (\vect{\psi}^*,\vect{\psi} )	&=  \prod\limits^{ \mathcal{N} }_{n=1} \rho^{n}_{\mathcal{W}} (  \psi^*_n, \psi_n  ), \\
\rho^{n}_{\mathcal{W}} (  \psi^*_n, \psi_n  ) & = 
\left\lbrace
\begin{array}{ll}
 \frac{ 2 }{ \pi } e^{-2|\psi_n|^2} (4 \psi^{*}_n \psi_n - 1) 	& {\rm if} \hspace{1.0mm} n=1 \\
 \frac{ 2 }{ \pi } e^{-2|\psi_n|^2}								& {\rm if} \hspace{1.0mm} n \neq 1 
\end{array}
\right. .
\end{align}
\endgroup
The pseudo probability density function $\rho^1_{\mathcal{W}}$ can be represented as a product of $\rho^{1,abs}_{\mathcal{W}}$, ${\rm Sign}( \rho^1_{\mathcal{W}}, \psi^*_1, \psi_1 )$, $N(\rho^1_{\mathcal{W}} )$ in the same way as explained in Sec. \ref{subsec:Wigner_formalism}.

\begin{figure}
	\centering
	\includegraphics[width=\linewidth]{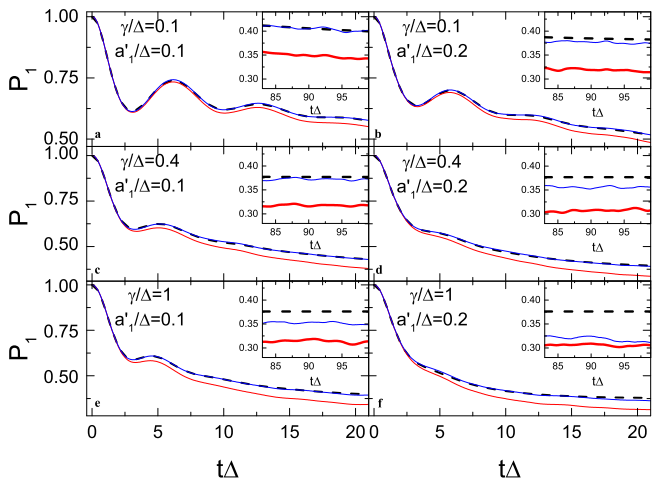}
	\caption{ (Color online) Population $P^{}_1$ of the first site for $\Omega = a'_2 = 0$. Exact solution (black dashed), TWA (red thick) and CTWA (blue thin). Inset: long time behavior of the same observable. 
	}
	\label{fig:Graph_1}
\end{figure}

\begin{figure}
	\centering
	\includegraphics[width=\linewidth]{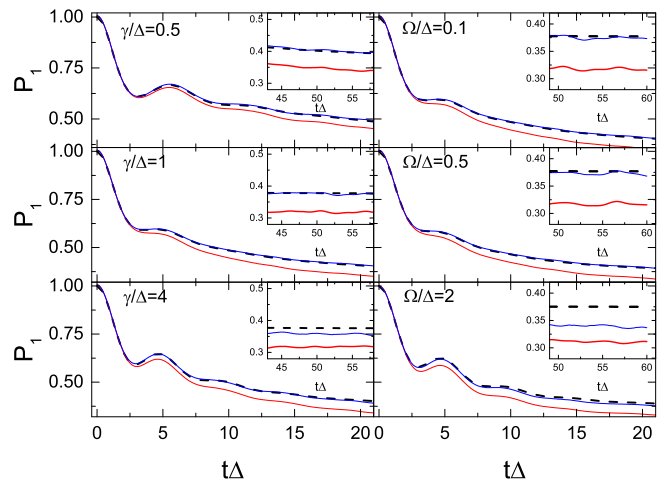}
	\caption{	(Color online) Population $P^{}_1$ of the first site for  $a'_2/ \Delta=0.1, a'_1/ \Delta=0 $. Left column: $\Omega/ \Delta = 0.1$. Right column: $\gamma/\Delta =1$. Exact solution (black dashed), TWA (red thick) and CTWA (blue thin).  Inset: long time behaviour of the same observable.	}
	\label{fig:Graph_2}
\end{figure}

	Information about the quality of the proposed approximation can be obtained from Fig. \ref{fig:Graph_1},\ref{fig:Graph_2} where we have compared our results with those from the TWA and with the results obtained by the use of the HEOM method which will be referred to as the exact results. In all figures we plot the population $P^{}_1$ of the first site. To understand why our approximation works better in some regimes, we will look at the elements of ${\rm diag}[ \vect{\varepsilon}^{1T} \vect{\phi}^1(t),	\vect{\varepsilon}^{2T} \vect{\phi}^2(t) ]$, that are contained in $\tilde{h}(t)$, defined in \eqref{eq:Langevin_Semiclass_3}. We can replace $\vect{\phi}^n(t)$ with the solution of \eqref{eq:phi_continuous}, which is obtained from \eqref{eq:phi_path_solution} after replacing $n^o(\tau)$ with $2| \psi (\tau) |^2$. It follows that the diagonal matrix plays a role of a memory term, whose elements are given by ($n$-index is omitted):
\begingroup
\allowdisplaybreaks
\begin{align}
\vect{\epsilon}^{T}_{} \vect{\phi}(t) & =
\vect{\varepsilon}^{T}_{} \big[  e^{At}_{}\vect{\phi}(0) + 
		\int^{t}_{0} e^{A(t-\tau)}_{} \vect{v} 2|\psi(\tau)|^2 d\tau \nonumber \\
 & + 	\int^{t}_{0} e^{A(t-\tau)}_{} B d\vect{W}(\tau)
 \big]  \label{eq:Some_Def_line_1}  \\
 & =  \vect{\varepsilon}^T_{} e^{At}_{} \vect{\phi}(0) + 2 a' \int^{t}_{0} g(t-\tau) |\psi(\tau)|^2_{} d\tau  \nonumber \\ 
 & +  \sqrt{a'} \int^{t}_{0} g(t-\tau) \mu dW_1(\tau),  \label{eq:Some_Def_line_2}
\end{align}
\endgroup
where $a'$, $\mu$ and $g(t)$ are given by
\begin{align}
a' & =
\left\lbrace
\begin{array}{ll}
a'_2 & \hspace{31.0mm} a'^{}_1 = 0 \\
a'_1 & \hspace{31.0mm}  \Omega = 0 
\end{array}
\right. , \label{eq:Def_ap} \\
\mu & =
\left\lbrace
\begin{array}{ll}
\sqrt{8T\gamma/(\gamma^2_{} \pm \Omega^2_{} )} & \hspace{8.6mm} a'^{}_1 = 0 \\
\sqrt{4T\gamma/\gamma^2_{}} & \hspace{8.6mm}  \Omega = 0 
\end{array}
\right. , \label{eq:Def_mu} \\
g(t) & =
\left\lbrace
\begin{array}{ll}
e^{-\gamma t }_{} S(\Omega t) ( \gamma^2_{} \pm \Omega^2_{} )/\Omega   & a'^{}_1 = 0 \\
e^{-\gamma t }_{} \gamma & \Omega = 0
\end{array}
\right.
. \label{eq:Def_gt}
\end{align}
The second term of \eqref{eq:Some_Def_line_2} is obtained by making use of the fact that the vectors $\vect{v},\vect{\varepsilon}$ are defined such that \eqref{eq:Eq_coefficients} is fulfilled. The equivalence $\mathcal{D}(t)/\pi = a' g(t)$ can be seen by a direct substitution of \eqref{eq:New_coupl_const}, \eqref{eq:Diss_part_example} into \eqref{eq:Some_Def_line_1}. The proof of $ \vect{\varepsilon}^Te^{At}Bd\vect{W}(t) = (a')^{1/2} g(t) \mu dW_1(t)$ is given in the Appendix \ref{sec:App_details_DA_model}.
	
	We are mainly interested in the contributions to \eqref{eq:Some_Def_line_2} containing $g(t)$ since they are responsible for the impact of the $ \vect{\kappa}\vect{\kappa}^{*T}$-dependent terms on the time evolution of the system. From the property $\int^{\infty}_{0} g(\tau) d\tau = 1 $ it follows, that for the two different spectral densities given in \eqref{eq:Spec_fct_both}, the strength of the memory kernel $g(t)$ is the same but the way how the previous values of $\psi(t)$, $W(t)$ are taken into account is different.
	
	The common feature of all cases, where the approximation is accurate at large time scales, is that the weight of the $g(t)$ function is uniformly distributed over large time scales and it does not change sign. The measure, that defines if a time interval is sufficiently large, is determined by the time scale $\tau_S = 1/h_{max}$, where $h_{max}$ is the maximum absolute value of the matrix representation of the commutator $H^{\times}_S : X \mapsto [h,X]$. For our system Hamiltonian it follows that $\tau_S = 1/\Delta$.
	
	If we consider the case, where $\Omega=0$ for the spectral densities of both environments, then the condition, that $g(t)$ is uniformly distributed over a large time interval is equivalent to $\tau_S \ll 1/\gamma \Leftrightarrow \Delta \gg \gamma $. This can be seen in Fig. \ref{fig:Graph_1}, where an increase of $\gamma/\Delta$ from $0.1$ to $1$ decreases the quality of the approximation at large time scales ($t\Delta = 100$). 
	
	If we consider the case $a'_1=0$ and $(C,S)=( \cos , \sin )$ for the spectral functions of both environments, then the condition for $g(t)$ is equivalent to $h_{max}> \gamma> \Omega$, where the second inequality comes from the restriction that $g(t)$ does not have to change sign over the $\tau_S$-timescale. The change of the quality of the approximation by changing $\gamma,\Omega$ of both spectral densities can be seen in Fig.\ref{fig:Graph_2}. In the first (second) column of the figure we see how an increase of $\gamma$ ($\Omega$) leads to a decrease of the quality of the approximation because the condition $h_{max}>\gamma $ ($\gamma>\Omega$) is violated. 
	
	The common feature of all figures is that our approximation reproduces more accurate results at large time scales than the TWA does, as long as the condition for the form of the memory kernel $g(t)$ is fulfilled and the system bath coupling $a'$ is sufficiently small. At larger couplings as well as for parameters of the environment that violate the condition for the memory kernel the quality of the approximation decreases as it is shown in the second column of Fig. \ref{fig:Graph_1}. Similarly to the TWA, the time range of validity of our approximation increases by an increase of the temperature.

\begin{figure}
	\centering
	\includegraphics[width=\linewidth]{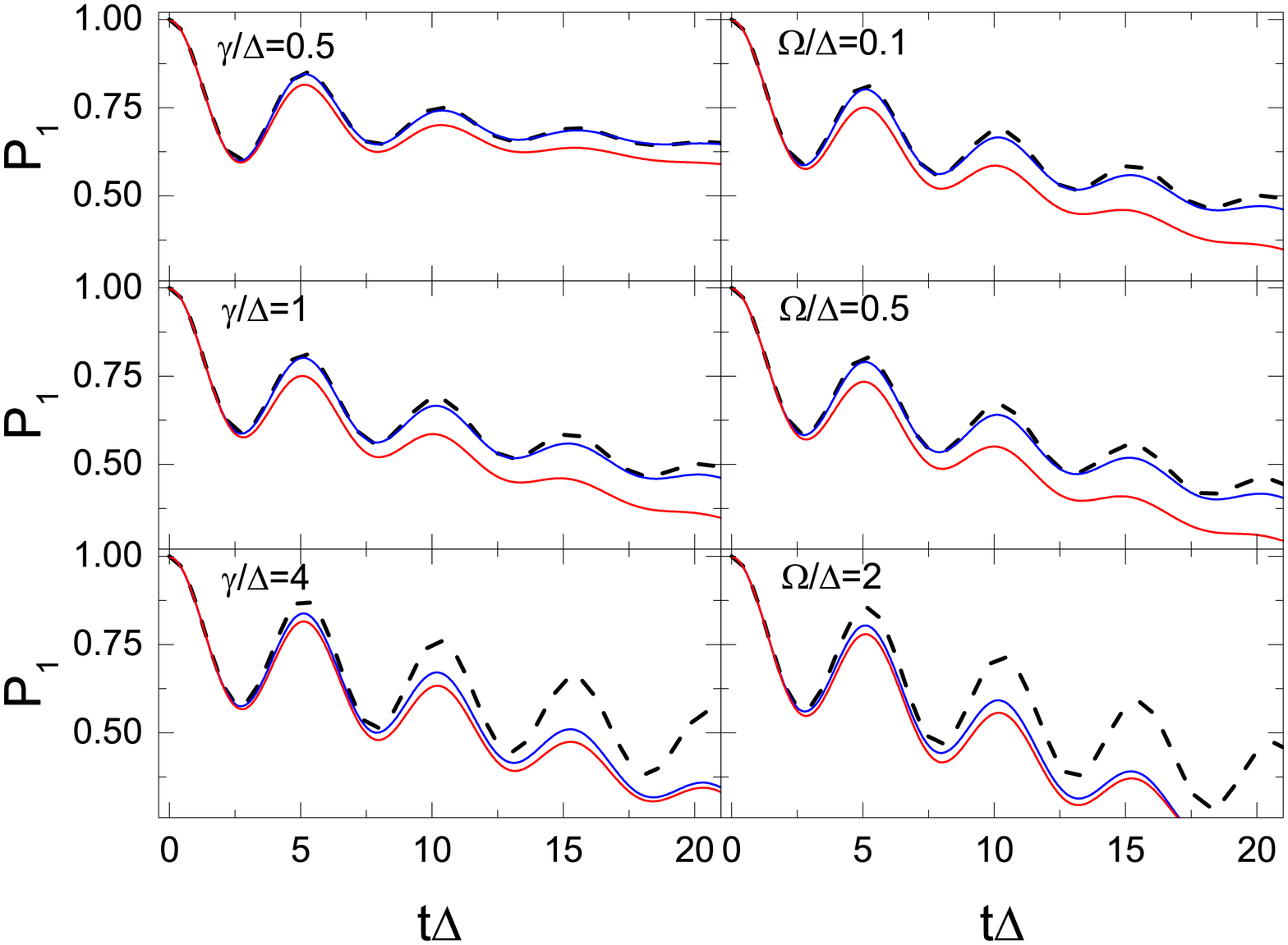}
	\caption{	(Color online) Population $P^{}_1$ of the first site for  $a'_2/ \Delta=0.1, a'_1/ \Delta=0 $. Left column: $\Omega/ \Delta = 0.1$. Right column: $\gamma/\Delta =1$. Exact solution (black dashed), TWA (red thick) and CTWA (blue thin). The temperature is decreased from $T/\Delta = 2$ (Fig. \ref{fig:Graph_2}) to $T/ \Delta = 0.2$. }
	\label{fig:Graph_3}
\end{figure}

	To test the new approximation at low temperatures we have considered again the case $a'_1=0$, $(C,S)= (\cos, \sin )$.
The time evolution of the population of the donor site is plotted in Fig. \ref{fig:Graph_3} for the same six cases as those considered in Fig. \ref{fig:Graph_2} but with temperature decreased from $T/\Delta =2 $ to $T/\Delta =0.2 $. We see that the CTWA still behaves better than the TWA but, as expected, both approximations fail to describe the short time behavior of the system accurately. The main reason is that the relative weight between the temperature dependent noise kernel $ \mathcal{F}(t)$ of the action that can always be taken exactly into account by the use of stochastic methods, and the temperature independent dissipation kernel $ \mathcal{D}(t)$ whose effects are always approximated in the TWA and CTWA, decrease. In this case the noise produced by $ \mathcal{F}(t)$ is not strong enough to overcompensate the effects from the dissipation kernel after applying the CTWA.

\section{Summary}
\label{sec:Conclusions}

	In this paper we have derived an alternative method to describe the exciton transport in open quantum systems. Instead of trying to derive an exact equation of motion for the reduced density matrix of the system, $\hat{\rho}_S$, we have used the fact that $\hat{\rho}_S $ lies completely in the single exciton subspace which allowed us to represent the time evolution of every non-zero element of $\hat{\rho}_S$ as the expectation value of a system operator, as shown in \eqref{eq:Def_rho_1ex_space},\eqref{eq:Def_rho_el_1ex_space}. This difference led also to a change of the requirements for the form of the effective action of the system that is obtained after applying some stochastic unraveling approach to it. Instead of having to derive a time-local action at the cost of introducing integrals over new variables that can later be interpreted as Gaussian random variables, and trying to obtain a time-local stochastic equation of motion for $\hat{\rho}_S$ \cite{stockburger2002exact,imai2015application,Stockburger_1999_n,Stockburger_2001_n}, we have derived an action that is completely real and linear in the quantum fields at the cost of introducing integrals over new variables that are weighted with pseudo-probability density functions. Even though in this case one can derive an exact equation of motion for the classical variables this method is numerically expensive. The rapid increase of the number of trajectories that are needed for the calculation of some observable origins from the new pseudo-probability density functions, which can modify the weight and the total sign of every contribution to the observable. 
	
	To overcome this problem we have derived an approximation of the equation of motion of the classical variables, where only some moments of the new variables weighted with the corresponding pseudo-probability density functions appear. We have shown that the quality of the approximation in the weak coupling limit depends on the form of the dissipation kernel of the Feynman-Vernon influence functional describing the effect of the environment on the system. If this kernel dissipates slowly in time without changing its sign, the corrections to the TWA give exact results even at large time scales ($t/\tau_S \gg 1$). This makes our approach applicable in situations, where the slow decay of the dissipation kernels does not allow the application of the Markovian approximations to the environment. 
	
	The method still remains simple enough to apply it to large systems. For a system composed of $\mathcal{N}$ sites, where each of them is linearly coupled to a separate environment, we can describe the time evolution of the system density matrix by a time local equation of an $\mathcal{N} \times \mathcal{N}$ matrix and $\mathcal{N}$ time local equations of vectors, where every vector describes the effect of one of the environments on the system. The dimension of each of those vectors is equal to the number of exponentially decaying functions, whose linear combination can describe the noise and dissipation kernel part of the Feynman-Vernon functional of the corresponding environment. In the limit of large networks the complexity of the approach depends mainly on the complexity of the time local equation of motion of the $\mathcal{N}\times \mathcal{N}$ matrix. It follows that the numerical effort to calculate a single trajectory of the observable is numerically as expensive as the calculation of some Markovian master equation in Lindblad form of the same system. So the increase of the numerical effort in comparison to the master equation depends on the number of trajectories of the observable which is needed to obtain a good approximation of its mean value. This number is of the order of $10^4-10^5$ and is required to sample correctly the initial state of the system which is described by a pseudo-probability density function. This is the price that we have to pay to take into account the non-Markovian effects of the environment on the system.

	\acknowledgments

We acknowledge support from the European Union (EU) through the Collaborative Project QuProCS (Grant Agreement No. 641277).

\appendix
\section{Proof of the mapping of the action to a time local expression}
\label{sec:App_Proof_1}
We have to show that
\begin{widetext}
\begin{align}
\label{eq:Eq_toshow_1}
& \exp \Big[ 
  \sum\limits^{}_{\tau} {\scriptstyle \Delta} t \hspace{1.0mm} i n^{\times}_{\tau}	\sum\limits^{}_{\tau'<\tau}  {\scriptstyle \Delta}t \Big(
\sum\limits_{l \in L \cup \tilde{L} } \alpha^{F}_l e^{ \lambda_l(\tau-\tau') } in^{\times}_{\tau'} +
\sum\limits_{l \in L} \alpha^{D}_l e^{ \lambda_l (\tau-\tau') } n^{o}_{\tau'}
\Big) \Big] \nonumber \\
 & = \int D[\vect{x}] D[\vect{\phi}] D[\vect{\varphi}] \exp \Big[ 
\sum\limits^{}_{\tau} i 2 \vect{\varphi}^{T}_{\tau}  
\Big( - \vect{\phi}_{\tau+\Delta t} + \vect{\phi}_{\tau}	+ {\scriptstyle \Delta}t A \vect{\phi}_{\tau}	+{\scriptstyle \Delta}t \vect{v} n^{o}_{\tau}	+ \sqrt{ {\scriptstyle \Delta} t}  B\vect{x}_{\tau}	\big)
+ \sum\limits^{}_{\tau} {\scriptstyle \Delta}t i n^{\times}_{\tau} \vect{\varepsilon}^{T}_{} \vect{\phi}_{\tau}
\Big] \rho_{\Phi} (\vect{\phi}_0 ),
\end{align}
\end{widetext}
where $\rho$ and $\Sigma$ are defined in \eqref{eq:Initial_distribution}, \eqref{eq:Initial_distribution_Sigma} and the number of elements in $L \cup \tilde{L}$ is $\mathcal{M}$. To do this we integrate out all $\varphi$-fields by the use of \eqref{eq:Eq_Delta_fct}. The equation of motion for the $\phi$-fields, defined by the set of $\delta$-functions, and its solution are given by:
\begin{align}
 d\vect{\phi}(t)	& = \big(  A\vect{\phi}(t) + \vect{v}n^{o}(t)  \big) dt + B d\vect{W}(t), \\
 \vect{\phi} (t)	& = e^{At}_{} \vect{\phi}(0) + \int^{t}_{0} e^{A(t-\tau)}_{} \vect{v}n^{o} (\tau) d\tau \nonumber \\
 & + \int^{t}_{0}  e^{A(t-\tau)}_{}Bd\vect{W} (\tau). \label{eq:phi_path_solution}
\end{align}
We replace $\vect{\phi}_{\tau}$ in the last term of the second line of \eqref{eq:Eq_toshow_1}  with the solution in \eqref{eq:phi_path_solution}. In the next step we integrate out the $x_{m,\tau}$ variables, which requires the replacement $\int d\vect{W} (\tau) = \sum_{\tau} \sqrt{ {\scriptstyle \Delta} t } \vect{x}_{\tau} $. This gives the following identity
\begin{align}
& \int D[\vect{x}] \times \nonumber \\
& \times \exp \bigg[
i \int^{t}_{0} d\tau \int^{\tau}_{0} d\tilde{\tau} 
n^{\times} (\tau) \vect{\varepsilon}^{T}_{} e^{A(\tau -\tilde{\tau})}_{} B d \vect{W} (\tilde{\tau})
\bigg] \nonumber \\
& = \exp \bigg[ 
-\frac{1}{2} \sum\limits^{ }_{m,\tilde{\tau}}  \bigg(	\sqrt{ {\scriptstyle \Delta}t }	\int\limits^{t}_{\tilde{\tau}} d\tau  n^{\times} (\tau) \vect{\varepsilon}^{T} e^{A( \tau-\tilde{\tau} )} B	\bigg)^2_m
\bigg] \nonumber \\
& = \exp \bigg[ 
 \int\limits^{t}_{0}	d\tau	in^{\times} (\tau)	\int\limits^{\tau}_{0}	d\tau'	in^{\times} (\tau')	m (\tau,\tau')
\bigg].
\end{align}
where $m \in \lbrace 1, \ldots \mathcal{M} \rbrace$ and the function $m(t,t')$ is defined as:
\begin{align}
m (t,t') & \overset{ t>t'}{=}  \int\limits^{ t' }_{0} d\tau \vect{\varepsilon}^{T}_{} e^{ A( t - \tau) }_{} B B^{T}_{} e^{ A^{T}_{}(t' - \tau) }_{} \vect{\varepsilon} .
\end{align}
The second line of \eqref{eq:Eq_toshow_1} transforms to:
\begin{align}
& \int \prod\limits^{\mathcal{M}}_{m=1} d\phi_{m,0} \exp \Big[
i \int^{t}_{0} d\tau  n^{ \times } (\tau) \vect{\varepsilon}^{T}_{} e^{ A \tau }_{} \vect{ \phi }^{}_{0} \Big]
\rho (\vect{\phi}^{}_{0}) \nonumber \\
& \times \exp \Big[
 \int^{t}_{0}  d\tau i n^{ \times } (\tau)  \int^{ \tau }_{ 0 } d\tau' 
\vect{\varepsilon}^{T} e^{ A (\tau - \tau') } \vect{v} n^{o} (\tau')		\Big] \nonumber \\
& \times \exp \Big[
\int^{t}_{0} d\tau i n^{ \times } (\tau)  \int^{\tau}_{0}  d\tau' m(\tau,\tau') i n^{\times} (\tau')
\Big] \label{eq:Appendix_eq_1}
\end{align}

	In the next step we have to decompose $m(t,t')$ into a part that does and a part that does not depend on $(t-t')$, where the first part has to be identified with the noise kernel of the action. To do this we choose the matrix $A$ such that the set of its eigenvalues coincides with $\lbrace 	\lambda_l	\rbrace_{l\in L \cup \tilde{L}} $. We assume, that each of the eigenvalues of $A$ has an algebraric and geometric multiplicity of one. It follows, that there exists an invertible matrix $S$ and a diagonal matrix $D$  ($S,D\in \mathbb{R}^{\mathcal{M}\times \mathcal{M} }$) such that $S^{-1}DS=A$. By the use of the definitions
\begin{align}
\tilde{\vect{\varepsilon}} & = (S^{-1})^T \vect{\varepsilon}, \\
F			& = SBB^TS^T, \\
\tilde{D}( f (\vect{\lambda}) )	& = {\rm diag} [	f (\lambda_1), \ldots ,	f (\lambda_{\mathcal{M}})  ], \\
\tilde{F}_{kl} (t)	& = -F^{}_{kl} \frac{ e^{\lambda_{k}t } }{ \lambda_k + \lambda_l }, \hspace{5.0mm} k,l \in \lbrace 1 \ldots \mathcal{M} \rbrace
\end{align}
we can show, that
\begingroup
\allowdisplaybreaks
\begin{align}
& \int^{t'}_{0} d\tau \big[ \tilde{D}( e^{\lambda(t-\tau)}_{} ) F \tilde{D}( e^{\lambda(t'-\tau)}_{} )   \big]_{kl} \nonumber \\
& = -F_{kl} \frac{ e^{\lambda_k (t-t')}_{} }{\lambda_k + \lambda_l} + F_{kl}\frac{ e^{\lambda_k t + \lambda_l t'} }{ \lambda_k + \lambda_l } \nonumber \\
& = -F_{kl} \frac{ e^{\lambda_k (t-t')}_{} }{\lambda_k + \lambda_l} - F_{kl} e^{\lambda_k t + \lambda_l t'} \int^{\infty}_{0} d\tau e^{(\lambda_k + \lambda_l)\tau}_{} \nonumber \\
& = \tilde{F}_{kl} (t-t') 
- \int^{ \infty }_{0} d\tau \big[	\tilde{D}( e^{\lambda (t + \tau)} ) F \tilde{D} (e^{\lambda (t'+\tau)}_{})	\big]_{kl} .
\end{align}
\endgroup
This allows us to decompose $m(t,t')$ as a sum of the following functions:
\begin{align}
 m (t,t') & = \tilde{\vect{\varepsilon}}^T \tilde{F} (t,t') \tilde{\vect{\varepsilon}} \nonumber \\
& - \int^{\infty}_{0} d\tau \vect{\varepsilon}^T e^{A(t+\tau)}_{} BB^T e^{A^T( t'+\tau)}_{} \vect{\varepsilon} \nonumber \\
& = \tilde{\vect{\varepsilon}}^T \tilde{F} (t,t') \tilde{\vect{\varepsilon}} 
- \vect{\varepsilon}^T e^{A t}_{} \Sigma e^{A^T t'}_{} \vect{\varepsilon} . \label{eq:m_decomposition}
\end{align}
After integrating out the $\phi^{}_0$-fields, the first line of \eqref{eq:Appendix_eq_1} modifies to
\begin{align}
& \exp \Big[  -\int^{t}_{0} d\tau  \int^{\tau}_{0} d\tau' n^{\times}(\tau) n^{\times} (\tau') \vect{\varepsilon}^{T}_{} e^{ A\tau }_{} \Sigma e^{ A^{T}_{} \tau'}_{} \vect{\varepsilon} \Big],
\end{align}
which cancels with the second term of \eqref{eq:m_decomposition}. It follows, that \eqref{eq:Appendix_eq_1} transforms to 
\begin{align}
& \hspace{2.8mm} \exp \Big[ \int^{t}_{0} d\tau in^{\times} (\tau) \int^{\tau}_{0} d\tau' \vect{\tilde{\varepsilon}} \tilde{F}(\tau-\tau') \vect{\tilde{\varepsilon}} in^{\times} (\tau') \Big] \nonumber \\
& \times \exp \Big[ 
\int^{t}_{0} d\tau in^{\times} (\tau) \int^{\tau}_{0} d\tau' \vect{\varepsilon}^{T}_{} e^{A(\tau-\tau') }_{} \vect{v} n^{o} (\tau') 
\Big].
\end{align}
A direct comparison of the last equation with \eqref{eq:Eq_toshow_1} and use of the fact, that the eigenvalues of $A$ are non-degenerate leads to the following equation for $\vect{v},\vect{\varepsilon},b_1, \ldots b_{\mathcal{M}}$:
\begin{align}
\alpha^{D}_l	& = \vect{ \tilde{ \varepsilon } }^{}_l  (S \vect{v})^{}_l, \\
\alpha^{F}_l	& = \sum\limits^{}_{k} \vect{\tilde{\varepsilon}}^{}_l \frac{ F^{}_{lk} }{ - (\lambda^{}_l + \lambda^{}_{k} )}  \vect{\tilde{\varepsilon}}^{}_k \hspace{4.0mm} l\in\lbrace 1 \ldots \mathcal{M} \rbrace. 
\end{align}

\section{Proof of the linearization of the action in the quantum variables}
\label{sec:App_proof_linearize_action_2}

	To show that \eqref{eq:Wigner_Identity_2a} is fulfilled, we have to integrate out the $x^{n}_{m,\tau}$ variables in the right-hand side of the equation. If we isolate only the $x$-terms and the terms that are non-linear in $\eta$ from this expression, we will obtain the following identity for every $n \in \lbrace 1 , \ldots \mathcal{N} \rbrace $:
\begin{align}
& \int D[\vect{x}^n] \exp \Big[
\sum\limits^{}_{\tau} i2\varphi^{nT}_{\tau}( {\scriptstyle \Delta} t\vect{v}^n \frac{| \eta_{n,\tau}|^2}{2}  + \sqrt{ {\scriptstyle \Delta}t} B^n \vect{x}^n_{\tau})  \Big] \nonumber \\
& = \prod\limits^{}_{\tau} \prod\limits^{ \mathcal{M}_n }_{m=1}  \exp \Big[ -\frac{1}{2} {\scriptstyle \Delta}t 4 (b^{n}_{m})^2 ( \varphi^{n}_{m,\tau} )^2  \nonumber \\
		& \hspace{40.0mm}	+ i {\scriptstyle \Delta}t v^n_m |\eta_{n,\tau}|^2 \varphi^n_{m,\tau} \Big]. 
\end{align}
We will present a set of transformations, that have to be applied to every $(n,m,\tau)$ contribution to the last expression. For better legibility we will omit these indices. We introduce the initially unknown function $f(x,y,z)$, which has to fulfill the following equation:
\begin{align}
& \exp [ -\frac{1}{2}\alpha \varphi^2 \pm i\beta \varphi \Re \eta^2 \pm i \beta \varphi \Im \eta^2 ] = \nonumber \\
& \int dx dy dz f(x,y,z) \exp [ i\varphi x + i \Re\eta y+ i \Im \eta z ], \label{eq:New_pdf_ansatz} \\
& \alpha = {\scriptstyle \Delta}t 4b^2, \hspace{3.0mm} \beta = {\scriptstyle \Delta}t |v|, \nonumber
\end{align}
where the integrals over $x,y,z$ are from $-\infty$ to $\infty$. The function $f(x,y,z)$ can be obtained by an inverse Fourier transformation:
\begin{align}
f(x,y,z)& = \int \frac{d\varphi d\Re \eta d\Im \eta}{(2\pi)^3} 
\exp [ -i\varphi x - i \Re \eta y - i \Im \eta z ]	\nonumber \\
		& \times \exp [ -\frac{1}{2}\alpha \varphi^2 \pm i \beta \varphi \Re \eta^2 \pm i \beta \varphi \Im \eta^2 ].
\end{align}
We have to point out that $\alpha$ and $\beta$ are positive constants, which is a necessary condition to perform the following transformations. We can integrate out $\varphi$ and then apply the following variable transformations: $(\Re \eta, \Im \eta) \rightarrow (\Re \eta \beta^{-1/2} \alpha^{1/4}, \Im \eta \beta^{-1/2} \alpha^{1/4} ) $, $(\Re \eta, \Im \eta) \rightarrow ( \rho \cos (\varepsilon), \rho \sin (\varepsilon) ) $ with $\rho = |\eta|$. By the use of the relation $y \cos (\varepsilon) + z \sin (\varepsilon) = \sqrt{y^2+z^2} \sin (\varepsilon + \delta) $, where $\delta = \arcsin (y/\sqrt{y^2+z^2})$, we can integrate out the $\varepsilon$ variable. The function $f(x,y,z)$ is then equal to:
\begin{align}
f(x,y,z) & = \frac{ \exp [- \frac{1}{2} \frac{x^2}{\alpha}] }{\sqrt{2\pi \alpha}} 
\int\limits^{\infty }_{0} d\rho \rho J_0 \big(\rho \sqrt{y^2+z^2} \big) \nonumber \\
		& \times \frac{1}{2\pi \beta \alpha^{-1/2}} \exp \Big[ -\frac{1}{2} \Big( \rho^4 \mp 2\rho \frac{x}{\alpha^{1/2}} \Big) \Big] .
\end{align}
We insert the result back in \eqref{eq:New_pdf_ansatz} and apply the following variable transformations: $(x,y,z) \rightarrow $ $ (x \alpha^{1/2} ,y \beta^{1/2} \alpha^{-1/4} ,z\beta^{1/2} \alpha^{-1/4})$ and $(y,z) \rightarrow $  $ (r \cos (\theta), r \sin (\theta)) $. The right-hand side of \eqref{eq:New_pdf_ansatz} is then equal to:
\begin{align}
& \int dx d\theta dr f_X (x) f_{\Theta} (\theta) f_{R|X}(r|x) \nonumber \\
& \times \exp \Big[ \mp i \varphi x \alpha^{1/2} + i2\Re \big( \eta^* \frac{r}{2}e^{i\theta} \beta^{1/2} \alpha^{-1/4} \big) \Big], \label{eq:New_pdf_ansatz_line_6}
\end{align}
which completes the proof of \eqref{eq:Wigner_Identity_2a}. By comparing the left side of \eqref{eq:New_pdf_ansatz} and \eqref{eq:New_pdf_ansatz_line_6} we see, that a change of the sign of the $i\varphi x \alpha^{1/2}$ expression does not turn \eqref{eq:New_pdf_ansatz_line_6} into it its complex conjugate as it is the case for the left side of \eqref{eq:New_pdf_ansatz}. This is related to the fact, that only specific combinations of random variables have a nonzero expectation value.

\section{Details of the application of the mapping procedure}
\label{sec:App_details_DA_model}

	To apply the mapping described in \ref{subsec:Mapping_proc}, where the noise and dissipation kernels of $i\mathcal{S}^n_{B,SB}$ are given by \eqref{eq:Diss_part_example},\eqref{eq:Def_F_wieder} and the sum over $l$ in \eqref{eq:Def_F_wieder},\eqref{eq:Def_fS},\eqref{eq:Def_fA} is neglected, we use the matrix ($n$-superscript will be omitted)
\begin{align}
A & = 
\left[
\begin{array}{cc}
0									&	( \gamma^2 \pm \Omega^2 )^{1/2} \\
-( \gamma^2 \pm \Omega^2 )^{1/2} 	&	-2\gamma
\end{array}
\right],
\end{align}
which has the same eigenvalues $\lambda_{1,2}$ as in \eqref{eq:Eigenvalues}. The equations for the coefficients of the vectors $\vect{v},\vect{\varepsilon} \in \mathbb{R}^2$ and for the diagonal elements of the matrix $B={\rm diag} [b_1,b_2]$ that were derived in Appendix \ref{sec:App_Proof_1}, are equivalent to the following set of equations:
\begingroup
\allowdisplaybreaks
\begin{widetext}
\begin{align}
(\varepsilon_1 v_1 - \varepsilon_2 v_2)\gamma + (\varepsilon_1 v_2 - \varepsilon_2 v_1)( \gamma^2 \pm \Omega^2 )^{1/2} & = a_2 \Omega,  \label{eq:Some_Eq_line_1}\\
\varepsilon_1 v_1 + \varepsilon_2 v_2 & = a_1, \label{eq:Some_Eq_line_2} \\
(b_2 \varepsilon_2)^2 + (b_1 \varepsilon_1)^2	& = 4T[ \gamma(a_1 f_S + a_2 f_A) - \Omega(a_2 f_S \mp a_1 f_A ) ] \nonumber \\ & = 4Ta_1, \label{eq:Some_Eq_line_3} \\
(\varepsilon_1 b_2)^2(\gamma^2 \pm \Omega^2) + [ 2\gamma(\varepsilon_1 b_1) - (\varepsilon_2 b_1)(\gamma^2 \pm \Omega^2)^{1/2} ]^2
& = 4T[ \gamma(a_1 f_S + a_2 f_A) + \Omega(a_2 f_S \mp a_1 f_A ) ](\gamma^2 \pm \Omega^2) \nonumber \\ & = 4T[ (\gamma^2 \pm \Omega^2)a_1 + 2\gamma \Omega a_2 ] . \label{eq:Some_Eq_line_4}
\end{align}
\end{widetext}
\endgroup
The first two equations can be derived from \eqref{eq:Eq_coefficients} by using 
\begin{align}
e^{At}	& = \frac{e^{-\gamma t}}{\Omega} \big[	\gamma S(\Omega t) \sigma_z + \Omega C(\Omega t)\mathbbm{1} + i \sqrt{\gamma^2 \pm \Omega^2} S(\Omega t) \sigma_y  \big]. \nonumber
\end{align}
The last two equations origin from the constraint that the part of $m(t,t')$ in \eqref{eq:Eq_coefficients_2} depending on the difference of the two arguments, is equal to $\mathcal{F}(t-t')$ which is given in \eqref{eq:Def_F_wieder}. In the second line of \eqref{eq:Some_Eq_line_3}, \eqref{eq:Some_Eq_line_4} we have used the assumption that we work at high temperatures and we have neglected all sums over $l$ in \eqref{eq:Def_F_wieder}, \eqref{eq:Def_fS}, \eqref{eq:Def_fA}.

	For the case $a'_a=0=a_1$ we can solve the four equations in the high temperature limit by setting $\varepsilon_1=b_2=v_2=0$ and for the case $\Omega=0$ we can set $b_2=v_2=0$. A direct calculation shows that $\vect{\varepsilon}^Te^{At}B d\vect{W}(t)= (a')^{1/2} g(t) \mu dW_1(t)$ where $a',g(t),\mu $ are defined in \eqref{eq:Def_ap}, \eqref{eq:Def_mu}, \eqref{eq:Def_gt}.

	The initial distribution of the $\phi^n_{m,0}$-variables, described by $\rho^n_{\Phi} (\vect{\phi}^n_0)$, is given in \eqref{eq:Initial_distribution} where the covariance matrix $\Sigma^n$, defined in \eqref{eq:Initial_distribution_Sigma}, takes the following form ($n$-index is omitted):
\begin{align}
\Sigma	& = \left[
\begin{array}{ccc}
\frac{ {\textstyle  b^2_1 + b^2_2 } }{ {\textstyle 4\gamma } } + \frac{ {\textstyle b^2_1 \gamma } }{ {\textstyle \gamma^2 \pm \Omega^2 } }	& \hspace{3.0mm} &	-\frac{ {\textstyle b^2_1 } }{ {\textstyle 2\sqrt{\gamma^2 \pm \Omega^2} } } \\
-\frac{ {\textstyle b^2_1} }{ {\textstyle 2\sqrt{\gamma^2 \pm \Omega^2} } }			& \hspace{3.0mm} & \frac{ {\textstyle b^2_1 + b^2_2 } }{ {\textstyle 4\gamma} }
\end{array}
\right].
\end{align}

\section{Alternative derivation of the CTWA and calculation of multitime correlation functions}
\label{sec:App_alt_deriv_CTWA}

	For the following discussion it will be useful to include the terms 
\begingroup
\allowdisplaybreaks
\begin{align}
&  \int d\tau \big( - \vect{\eta}^{*T}(\tau) \vect{\nu} (\tau) + \vect{\eta}^{T}(\tau) \vect{\nu}^* (\tau)   \big), \label{eq:Source_fields} \\
&  \vect{\nu}(t) = [\nu_1 (t), \ldots \nu_{\mathcal{N}} (t) ]^T  \nonumber
\end{align}	
\endgroup
in $i\mathcal{S} + \sum_n i\mathcal{S}^n_{B,SB} $ which will modify Eq. \eqref{eq:Path_psi_exact} by adding $ -\vect{\nu}_t {\scriptstyle \Delta} t $ on its right side. We can use the formal solution of this equation
\begin{align}
\vect{\psi}(t)		 & = \vect{\psi}_{TWA}(t) + \vect{\psi}_{QC}(t) + \vect{\psi}_{\nu}(t), \label{eq:Formal_solution} \\
\vect{\psi}_{TWA}(t) & = \hat{T} \exp \Big[ -i \int^t_{0} \tilde{h} (s) ds \Big] \vect{\psi}(0) , \nonumber \\
\vect{\psi}_{QC}(t) & = \sum\limits^{\mathcal{N}}_{n=1} \sum\limits^{ \mathcal{M}_n }_{m=1} \hat{T} \int^t_{0} 
\exp \Big[ -i \int^t_{\tau} \tilde{h} (s) ds \Big] \times  \nonumber \\
	&  \hspace{25.0mm} \times \vect{\kappa}^{nm} d\chi^n_{m}(\tau), \nonumber \\
\vect{\psi}_{\nu}(t)	 & = - \hat{T} \int^t_{0}  \exp \Big[ -i \int^t_{\tau} \tilde{h} (s) ds \Big] \vect{\nu}(\tau) d\tau , \nonumber \\
\int^t_{t'} f(\tau) d\chi^n_m (\tau)	& = \lim_{ {\scriptstyle \Delta}t \rightarrow 0 } \sum\limits^{}_{t' \leq \tau \leq t} f(\tau ) {\scriptstyle \Delta} \chi^n_{m,\tau} \nonumber
\end{align}
and construct $\vect{\psi} (t) \vect{\psi}^{*T}(t)$ for the case $\vect{\nu}(t) = \vect{\nu}^{*T}(t) = 0$. We can take the average over all $r^{n}_{m,\tau}$, $\theta^n_{m,\tau}$ variables ( denoted by $\langle \ldots \rangle_{r, \theta} $) and neglect the mixed terms $\vect{ \psi }_{TWA} (t) \vect{ \psi }^{ *T }_{QC} (t) $, $\vect{ \psi }_{QC} (t) \vect{ \psi }^{*T}_{TWA} (t) $. Then we can apply the following self-consistent approximation:
\begin{widetext}
\begin{align}
 \langle  \vect{\psi} (t) \vect{\psi}^{*T} (t) \rangle_{r,\theta} & = 
 \hat{T} \exp \Big[ -i\int^t_0 \tilde{h}(s) ds \Big] \vect{\psi} (0) \vect{\psi}^{*T} (0)
 \hat{T}^{\dagger} \exp \Big[ i\int^t_0 \tilde{h}(s) ds \Big] \nonumber \\
& + \sum\limits_{n,n'} \sum\limits_{m,m'} \int^t_0 \int^t_0 
\hat{T} \exp \Big[ -i\int^t_{\tau} \tilde{h}(s) ds \Big] \vect{\kappa}^{nm} \vect{\kappa}^{n'm' *T}
\hat{T}^{\dagger} \exp \Big[ i\int^t_{\tau'} \tilde{h}(s) ds \Big]  
\langle d\chi^n_m (\tau) d\chi^{n'}_{m'} (\tau')	\rangle_{r,\theta} \nonumber \\
 & = 
 \hat{T} \exp \Big[ -i\int^t_0 \tilde{h}(s) ds \Big] \vect{\psi} (0) \vect{\psi}^{*T} (0)
 \hat{T}^{\dagger} \exp \Big[ i\int^t_0 \tilde{h}(s) ds \Big] \nonumber \\
 & + 
 \sum\limits^{\mathcal{N}}_{n=1} \sum\limits^{\mathcal{M}_n}_{m} \int^t_0  
 \hat{T} \exp \Big[ -i\int^t_{\tau} \tilde{h}(s) ds \Big] \vect{\kappa}^{nm} \vect{\kappa}^{n'm' *T}
 \hat{T}^{\dagger} \exp \Big[ i\int^t_{\tau} \tilde{h}(s) ds \Big]  dW^n_m (\tau), \label{eq:Self_cons_approx}
\end{align}
\end{widetext}
where we have used \eqref{eq:Moment_A} to show the relation
\begin{align*}
\langle d\chi^n_m (\tau) d\chi^{*n'}_{m'} (\tau')	\rangle_{r,\theta} & \delta_{nn'} \delta_{mm'} \delta(\tau-\tau') dW^n_m (\tau).
\end{align*}
The $\tilde{h}$ matrix is defined in the same way as in \eqref{eq:Langevin_Semiclass_3} with the difference that the $\vect{\phi}$ terms are obtained from \eqref{eq:Langevin_Semiclass_2} by replacing $|\psi_{n,t}|^2$ term with the $nn$-component of $\langle \vect{\psi}(\tau) \vect{\psi}^{*T}(\tau) \rangle_{r,\theta }$ from \eqref{eq:Self_cons_approx}. The differential of \eqref{eq:Self_cons_approx} is then equal to \eqref{eq:psi_psi_dag_continuous}. 
	
	In the end we will briefly discuss the possibility to to use this approach to calculate multitime correlation functions. As example we will calculate the nonlinear response function $(t_3>t_2>t_1)$
\begin{align}
& {\rm tr} \big[ \hat{a}^{\textcolor{white}{\times}}_{n_3} (t_3) \hat{a}^{\dagger \times}_{n_2} (t_2)   \hat{a}^{\dagger \times}_{n_1} (t_1)  \hat{a}^{ \times }_{n_0} (t_0)  \hat{\rho}_{tot}  \big]  = \nonumber \\
& \int D[\vect{\psi},\vect{\eta}]  \psi^{ \textcolor{white}{*} }_{n_3,t_3} \eta^*_{n_2,t_2} \eta^*_{n_1,t_1}  \eta^{ \textcolor{white}{*} }_{n_0,t_0}	e^{i\mathcal{S}} \rho_{\mathcal{W}} ( \vect{\psi}^{*}_0, \vect{\psi}^{ \textcolor{white}{*} }_0 ) , \label{eq:Multitime_fct}
\end{align}
where $\hat{a}^{\times}_{n_j} (t) \hspace{0.5mm} \bullet  =  [ \hat{a}_{n_j} (t) , \bullet ] $ and $\hat{a}^{ \textcolor{white}{*} }_{n_j} (t)$ is an operator in the Heisenberg picture. To calculate \eqref{eq:Multitime_fct} we can use the idea proposed in \cite{polkovnikov_2010_phase}. We include again \eqref{eq:Source_fields} in the definition of the action and choose the $\vect{\nu} (t)$, $\vect{\nu}^*(t) $ to be equal to
\begingroup
\allowdisplaybreaks
\begin{align*}
\vect{\nu}(t)	& = \sum\limits^{2}_{j=0} \delta(t-t_j)  ({\scriptstyle \Delta}t)^y \nu_{n_j}, \\
\vect{\nu}^*(t)	& = \sum\limits^{2}_{j=0} \delta(t-t_j)  ({\scriptstyle \Delta}t)^y \nu^*_{n_j}, \hspace{3.0mm} y\in \mathbb{R}_+ \\
\end{align*}
\endgroup
and $ \eta^*_{n_2,t_2} \eta^*_{n_1,t_1}  \eta^{ \textcolor{white}{*} }_{n_0,t_0} $ is replaced with 
\begin{align*}
F	(\nu^*_{n_2}, \nu^{ \textcolor{white}{*} }_{n_2} ) 
F	(\nu^*_{n_1}, \nu^{ \textcolor{white}{*} }_{n_1} ) 
F^*	(\nu^*_{n_0}, \nu^{ \textcolor{white}{*} }_{n_0} ) ({\scriptstyle \Delta }t )^{-3y} , 
\end{align*}
where $F$ is defined such that:
\begingroup
\allowdisplaybreaks
\begin{align}
0 & = \int F(\nu^* , \nu) d\nu^* d\nu = \int F(\nu^* , \nu) \nu^* d\nu^* d\nu , \\
1 & = \int F(\nu^* , \nu) \nu d\nu^* d\nu.
\end{align}
\endgroup
The $F$-function can again be represented as a product of a real probability density function, a normalization factor and a ${\rm Sign}$-function, and the integration over the $\nu^{ \textcolor{white}{*} }_{n_j}$, $\nu^*_{n_j}$-variables can be carried in exactly the same way as the integration over over $\vect{\psi}_0$, $\vect{\psi}^*_0 $. A direct expansion of $e^{i \mathcal{S} }$ in powers of $\nu_j$, $\nu^*_j$ shows that in the limit ${\scriptstyle \Delta} t \rightarrow 0 $ the new and the old expression are equal. Our task now is to calculate the average of $\vect{\psi}_t$ over all $r^n_{m,\tau}$, $\theta^n_{m, \tau}$ variables. To do this we can use the formal solution \eqref{eq:Formal_solution} for $\vect{\psi}$ and neglect $\langle \vect{\psi}_{QC} (t) \rangle_{r,\theta}$ since $\langle  d\chi^n_{m,\tau} \rangle_{r,\theta} =0 $. The $| \psi_{n}(\tau) |^2$ term that appears in the diagonal elements of the $\tilde{h}(t)$-matrix can be replaced with the $nn$-term if $\langle \vect{\psi}(\tau) \vect{\psi}^{*T} (\tau) \rangle_{r,\theta} $ defined in \eqref{eq:Self_cons_approx}. If we calculate the differential of this approximation we will obtain the following equation (the $\langle \ldots \rangle_{r,\theta} $ brackets are neglected):
\begin{align}
\vect{\psi}_{t+ \Delta t} & = \vect{\psi}_t - i \tilde{h}(t) \vect{\psi}_t {\scriptstyle \Delta} t 
- \sum\limits^{2}_{j=0} \delta_{t,t_j} ({\scriptstyle \Delta}t)^y \nu_{n_j} .
\end{align} 
The only difference between the last equation and the corresponding TWA lies in the definition of the memory term in the diagonal elements of $\tilde{h}(t)$.

\bibliography{Main_text_incl_figures}
\end{document}